\documentclass{article}
\pdfoutput=1
\usepackage{lettrine}
\usepackage{epsfig}
\usepackage{graphicx}
\usepackage{color}
\usepackage{array}
\usepackage{pstricks}
\usepackage{titlesec}
\usepackage{makeidx}
\usepackage{amsmath}
\usepackage{bm}
\usepackage{mathrsfs}
\usepackage{amssymb}
\usepackage{makeidx}
\usepackage{pict2e}
\usepackage{supertabular}
\usepackage{fancyhdr}
\usepackage{pdfpages}
\usepackage{url}
\usepackage{natbib}
\usepackage{appendix}
\usepackage[a4paper,pdftex,verbose,margin=3.0cm,twoside]{geometry}
\newsavebox{\astrutbox}
\sbox{\astrutbox}{\rule[-5pt]{0pt}{20pt}}

\title{Small-scale dynamo action in rotating compressible convection}
\author{B. Favier\footnote{Corresponding author: benjamin.favier@ncl.ac.uk} and P.J. Bushby \\
School of Mathematics and Statistics, Newcastle University,\\ Newcastle upon Tyne NE1 7RU, UK
}
\date{}
\begin{document}
\maketitle

\abstract{
We study dynamo action in a convective layer of
electrically-conducting, compressible fluid, rotating about the
vertical axis. At the upper and lower bounding surfaces,
perfectly-conducting boundary conditions are adopted for the magnetic
field. Two different levels of thermal stratification are considered.
If the magnetic diffusivity is sufficiently small, the convection acts
as a small-scale dynamo. Using a definition for the magnetic Reynolds
number $R_M$ that is based upon the horizontal integral scale and the
horizontally-averaged velocity at the mid-layer of the domain, we find that rotation tends to reduce the critical
  value of $R_M$ above which dynamo action is observed. Increasing
  the level of thermal stratification within the layer does not
  significantly alter the critical value of $R_M$ in the rotating
  calculations, but it does lead to a reduction in this critical value
  in the non-rotating cases. At the highest computationally-accessible
  values of the magnetic Reynolds number, the saturation levels of
the dynamo are similar in all cases, with the mean magnetic energy
density somewhere between 4 and 9\% of the mean kinetic energy
density. To gain further insights into the
  differences between rotating and non-rotating convection, we
  quantify the stretching properties of each flow by measuring
  Lyapunov exponents. Away from the boundaries, the rate of
stretching due to the flow is much less dependent
  upon depth in the rotating cases than it is in the corresponding
  non-rotating calculations. It is also shown that the effects of
rotation significantly reduce the magnetic energy dissipation in the
lower part of the layer. We also investigate certain
  aspects of the saturation mechanism of the dynamo.  
}

%
%
\section{Introduction}

In a hydromagnetic dynamo, motions in an electrically-conducting fluid lead to the amplification of a seed magnetic field. 
Dynamo action can only occur if the inductive effects of the fluid
motions outweigh the dissipative effects of magnetic
diffusion. Provided that the magnetic energy density of the seed field
is very much less than the kinetic energy density of the flow, the
early stages of evolution of the dynamo process are effectively
kinematic, which implies that the magnetic energy in the system grows
exponentially with time (although the magnetic energy
  will tend to fluctuate about this exponential trajectory if the velocity
  field is time-dependent). Eventually, however, the nonlinear feedback of
the magnetic field upon the flow (via the Lorentz force) becomes
dynamically significant. This halts the growth of the dynamo-generated
magnetic field, leading to a saturated nonlinear dynamo. There are many examples of natural dynamos. For example, it is
generally accepted that the 11 year solar magnetic cycle \citep[see,
for example,][]{stix04}, is driven by an oscillatory dynamo. Similar
dynamo mechanisms drive cyclic magnetic activity on other stars
\citep{bal96}. Dynamo action is also responsible for sustaining the
Earth's magnetic field \citep[see, for example,][]{rob00}. It is also
possible to drive hydromagnetic dynamos in laboratories,  as
demonstrated by recent liquid metal experiments \citep[such as the
French VKS experiment, see for example,][]{mon09}.  

Convection plays a crucial role in many natural dynamos, particularly
in the context of solar, stellar and planetary dynamos. Although it is
possible to study global dynamo models numerically \citep{brun04},
most of the theoretical work on this problem has been based upon local
models of convection, where a fluid layer is heated from below and
cooled from above. In the context of Boussinesq convection, this
classical setup has been studied numerically by \citet{meneguzzi89}
and \citet{cattaneo99}. In highly-conducting fluids, Boussinesq
convection can act as an efficient dynamo, producing small-scale,
intermittent magnetic fields. In the Boussinesq approximation, the
effects of compressibility are neglected. More recent studies have
focused upon dynamo action in local models of convection in fully
compressible fluids \citep{vogler2007,brummell10,bushby10}. It is now
well-established that these compressible models can also drive
small-scale dynamos.  

Previous calculations have clearly demonstrated that rotation is not a
necessary ingredient for small-scale dynamo action in
convectively-driven flows. However, rotation is present in most
natural dynamos, and this additional physical effect can profoundly
influence the behaviour of a convective layer. For example, when the
rotation vector is aligned with the vertical axis, rapid rotation not
only inhibits convection, but it also reduces the preferred horizontal
scale of the convective instability \citep{chan61}. Even far from
onset, when compressible convection is fully turbulent, rotation can
also play an important role: helical convective plumes tend to become
aligned with the rotation vector. This is particularly apparent when
the axis of rotation is tilted away from the vertical
\citep{brummell96,brummell98a}. Given these hydrodynamical
considerations, we might reasonably expect the dynamo properties of
rotating convection to differ from the equivalent non-rotating
case. There have been numerous studies of this problem in Boussinesq
convection \citep{childress72,stpierre93,jones2000,stell04,cattaneo06}, but the
compressible case is less well understood. Existing studies have generally adopted complex models, with multiple polytropic layers,
inclined rotation vectors, or additional physical effects such as an
imposed velocity shear \citep[see, for
example,][]{brandenburg1996,kapyla2008,kapyla2009}. In fact, the
simpler problem of dynamo action in a single layer of turbulent
compressible convection, rotating about the vertical axis, is still
not fully understood. Therefore, the aim of this paper is to study the
ways in which rotation and compressibility influence the dynamo
properties of convection in a simple polytropic layer.  

The paper is organised as follows. The governing equations, boundary
conditions and parameters, together with the numerical methods are
discussed in the next section. Considerations about the dimensionless
numbers of interest, and the Lyapunov exponents of hydrodynamic convection are presented in Section
\ref{sec:hydro}. In Section \ref{sec:dynamo}, we
discuss results from a set of dynamo calculations. Finally, in
Section~\ref{sec:conclusions}, we present our conclusions.  

%
%
\section{Model and method}

\subsection{Model and governing equations}

We consider the evolution of a plane-parallel layer of compressible
fluid, bounded above and below by two impenetrable, stress-free walls,
a distance $d$ apart. The upper and lower boundaries are held at fixed
temperatures, $T_0$ and $T_0 + \Delta T$ respectively. Taking $\Delta
T>0$ implies that this layer is heated from below. The geometry of
this layer is defined by a Cartesian grid, with $x$ and $y$
corresponding to the horizontal coordinates. The $z$-axis points
vertically downwards, parallel to the constant gravitational
acceleration $\bm{g}=g\bm{e}_z$. The layer is rotating about the
$z$-axis, with a constant angular velocity
$\bm{\Omega}=\Omega\hat{\bm{z}}$. The horizontal size of the fluid
domain is defined by the aspect ratio $\lambda$ so that the fluid
occupies the domain $0<z<d$ and $0<x,y<\lambda d$. The physical
properties of the fluid, namely the specific heats $c_p$ and $c_v$,
the dynamic viscosity $\mu$, the thermal conductivity $K$, the magnetic permeability $\mu_0$ and the magnetic diffusivity $\eta$, are assumed to be constant. The model is identical to that used by \citet{matt95}, except for the addition of rotation.

It is convenient to introduce dimensionless variables, so we adopt the scalings described in \citet{bushby08}. Lengths are scaled with the depth of the layer $d$. The temperature $T$ and the density $\rho$ are scaled with their values at the upper surface, $T_0$ and $\rho_0$ respectively. The velocity $\bm{u}$ is scaled with the isothermal sound speed $\sqrt{R_*T_0}$ at the top of the layer, where $R_*=c_p-c_v$ is the gas constant. We adopt the same scaling for the Alfv\'en speed, which implies that 
the magnetic field $\bm{B}$ is scaled with $\sqrt{\mu_0\rho_0R_*T_0}$. Finally, we scale time by an acoustic time scale $d/\sqrt{R_*T_0}$. 

We now express the governing equations in terms of these dimensionless variables. The equation for conservation of mass is given by
\begin{equation}
\label{eq:mass}
\frac{\partial \rho}{\partial t}=-\nabla\cdot\left(\rho\bm{u}\right) \ .
\end{equation}
Similarly, the dimensionless momentum equation can be written in the following form, 
\begin{equation}
\label{eq:momentum}
\rho\left(\frac{\partial\bm{u}}{\partial t}+\kappa\sigma Ta_0^{1/2}\hat{\bm{z}}\times\bm{u}\right)=-\nabla\left(P+\frac{\bm{B}^2}{2}\right)+\theta\left(m+1\right)\rho\hat{\bm{z}}+\nabla\cdot\left(\bm{BB}-\rho\bm{uu}+\kappa\sigma\bm{\tau}\right) \ ,
\end{equation}
where $P$ is the pressure, given by the equation of state $P=\rho T$, and $\bm{\tau}$ is the stress tensor defined by
\begin{equation}
\tau_{ij}=\frac{\partial u_i}{\partial x_j}+\frac{\partial u_j}{\partial x_i}-\frac23\delta_{ij}\frac{\partial u_k}{\partial x_k} \ .
\end{equation}
Several non-dimensional parameters appear in Equation~\eqref{eq:momentum}. The parameter $\theta=\Delta T/T_0$ is the dimensionless temperature difference across the layer, whilst $m=gd/R_*\Delta T-1$  corresponds to the polytropic index.
The dimensionless thermal diffusivity is given by
$\kappa=K/d\rho_0c_p(R_*T_0)^{1/2}$ and $\sigma = \mu c_p/K$ is the
Prandtl number. Finally, $Ta_0$ is the standard Taylor number,
$Ta_0=4\rho_0^2\Omega^2d^4/\mu^2$, evaluated at the upper boundary. The induction equation for the magnetic field is
\begin{equation}
\label{eq:induction}
\frac{\partial \bm{B}}{\partial t}=\nabla\times\left(\bm{u}\times\bm{B}-\zeta_0\kappa\nabla\times\bm{B}\right) \ ,
\end{equation}
where $\zeta_0=\eta c_p\rho_0/K$ is the ratio of magnetic to thermal diffusivity at the top of the layer. The magnetic field is solenoidal so that
\begin{equation}
\nabla\cdot\bm{B}=0 \ .
\end{equation}
Finally, the heat equation is
\begin{equation}
\label{eq:heateq}
\frac{\partial T}{\partial t}=-\bm{u}\cdot\nabla T-(\gamma-1)T\nabla\cdot\bm{u}+\frac{\kappa\gamma}{\rho}\nabla^2T+\frac{\kappa(\gamma-1)}{\rho}\left(\sigma\tau^2/2+\zeta_0\left|\nabla\times\bm{B}\right|^2\right)  \ ,
\end{equation}
where $\gamma=c_p/c_v$.

To complete the specification of the model, some boundary conditions
must be imposed. In the horizontal directions, all variables are
assumed to be periodic. As has already been described, the
upper and lower boundaries are assumed to be impermeable and stress-free, which
implies that $u_{x,z}=u_{y,z}=u_z=0$ at $z=0$ (the upper boundary) and
$z=1$ (the lower boundary). Having non-dimensionalised the system, the
thermal boundary conditions at these surfaces correspond to fixing
$T=1$ at $z=0$ and $T=1+\theta$ at $z=1$. For the magnetic field
boundary conditions, two choices have typically been made in previous
studies. One can consider the upper and lower boundaries to be perfect
electrical conductors or one can adopt a vertical field boundary
condition. We choose appropriate conditions for perfectly-conducting
boundaries, which implies that $B_z = B_{x,z}=B_{y,z}=0$ at $z=0$ and
$z=1$. This is partly to facilitate comparison with the Boussinesq
study of \citet{cattaneo06}, but this is not the only motivation for
adopting boundary conditions that enforce $B_z=0$ at the upper
surface. Strong concentrations of vertical magnetic flux at the upper
surface tend to become partially evacuated \citep[see, for
example,][]{bushby08}, which dramatically increases the local Alfv\'en
speed (as well as significantly reducing the local thermal diffusion
time scale). This can impose very strong constraints upon the time-step
in any explicit numerical scheme. Hence, there are also numerical
reasons for adopting perfectly-conducting boundary conditions. In this
context, it is worth noting that it has been shown in the Boussinesq case \citep{thelen2000} that the dynamo efficiency of convection is largely
insensitive to the detailed choice of boundary conditions for the
magnetic field \citep[see also the compressible calculations
of][]{kapyla2010}. Although not reported here, we have carried out a
small number of simulations with vertical field boundary conditions,
and the results are qualitatively similar in that case. 

\subsection{Model parameters and initial conditions}

With the given boundary conditions, it is straightforward to show that these
governing equations have a simple equilibrium solution, corresponding
to a hydrostatic, polytropic layer:
\begin{equation}
\label{eq:polytrope}
T=1+\theta z\, , \; \rho = \left(1+\theta z\right)^m\,, \; \bm{u} = \bm{0}\,,\; \bm{B}=\bm{0}\ .
\end{equation}
This polytropic layer (coupled with a small thermal perturbation) is
an appropriate initial condition for simulations of hydrodynamic
convection. All the hydrodynamic flows that are considered in this
paper are obtained by evolving the governing equations from this basic
initial condition. Once a statistically-steady state has been reached, the dynamo properties of these flows can be investigated by inserting a weak (seed) magnetic field into the domain.  

There are many non-dimensional parameters in this system. Since
it is not viable to complete a systematic survey of the whole of
parameter space, we only vary a subset of the available
parameters. Throughout this paper, the polytropic index is fixed at
$m=1$, whilst the ratio of specific heats is given by $\gamma=5/3$ (the appropriate value for a monatomic
gas). These choices ensure that the initial polytropic layer is
unstable to convective perturbations. We also fix the Prandtl number
to be $\sigma=1$. In order to study the effects of varying the
stratification, we adopt two different values of $\theta$. The case of
$\theta=3$ corresponds to a weakly-stratified layer, whilst
$\theta=10$ represents a highly-stratified case in which the temperature and density both vary by an order of magnitude across the layer.

\begin{table}
 \begin{center}
\def~{\hphantom{0}}
 \begin{tabular}{cccccc}
   Run & $Ra$ & $Ta$ & $\theta$ & $\kappa$ & $Re$  \\
   $R1$ & $3\times10^5$ & $0$ & $3$ & $0.0055$ & $153$ \\
   $R2$ & $4.6\times10^5$ & $10^5$ & $3$ & $0.0044$ & $157$ \\
   $R3$ & $5\times10^5$ & $0$ & $10$ & $0.0242$ & $149$ \\
   $R4$ & $6\times10^5$ & $10^5$ & $10$ & $0.02$ & $152$ \\
  \end{tabular}
  \caption{The set of parameters for the four different cases. Note that the
    (mid-layer) Taylor number is defined by
    $Ta=4(1+\theta/2)^{2m}\rho_0^2\Omega^2d^4/\mu^2$. See the main
    part of the text for the definitions of the global Reynolds
    number, $Re$, and the Rayleigh number, $Ra$.\label{tab:one}} 
 \end{center}
\end{table}

The main aim of this study is to address the effects of
rotation and compressibility upon dynamo action in a convective
layer. So, for each value of $\theta$, we consider a rotating and a
non-rotating case (which implies four different cases overall).  Note
that some care is needed when comparing simulations of rotating
convection at different values of $\theta$. The Taylor number that
appears in the governing equations, $Ta_0$, corresponds to the Taylor
number at the top of the domain. Given the differing levels of
stratification, it makes more sense to specify the same \textit{mid-layer} Taylor number for each of the rotating cases. Since the Taylor number is proportional to $\rho^2$, the mid-layer Taylor number
(in the unperturbed polytrope) is defined by
$Ta=Ta_0(1+\theta/2)^{2m}$. Two different values of $Ta$ are
considered, $Ta=0$ (the non-rotating cases) and
$Ta=10^5$. Further
discussion regarding the depth-dependence of the Taylor number is
given in the next section. 

Another parameter that needs to be specified is the dimensionless
thermal diffusivity, $\kappa$. However, rather than prescribing values for
$\kappa$, we define the mid-layer Rayleigh number 
\begin{equation}
\label{eq:rayleigh}
Ra=\left(m+1-m\gamma\right)\left(1+\theta/2\right)^{2m-1}\frac{(m+1)\theta^2}{\kappa^2\gamma\sigma} \ ,
\end{equation}
which is inversely proportional to $\kappa^2$ and measures the
destabilising effect of the superadiabatic temperature gradient
relative to the stabilising effect of (viscous and thermal) diffusive processes. As described in the
Introduction, rotation tends to stabilise convection, whilst larger
values of $\theta$ also have a stabilising effect
\citep{gough76}. Therefore it does not make sense to keep $Ra$
constant in all cases. Instead, we vary the Rayleigh number from one
calculation to the other, ranging from $Ra=3\times10^5$ up to
$Ra=6\times10^5$. The aim was to reach similar values of the Reynolds
number in all cases. We shall discuss another possible definition for
the Reynolds number in the next section, but for now we define a
global Reynolds number, based upon the rms velocity ($U_{\textrm{rms}}$), the
kinematic viscosity at the mid-layer ($\kappa\sigma/\rho_{\textrm{mid}}$, where $\rho_{\textrm{mid}}$ is the mean density
at the mid-layer of the domain) and the depth of the layer (which
equals unity in these dimensionless units). Hence this global Reynolds
number is given by
\begin{equation}
Re= \frac{\rho_{\textrm{mid}}U_{\textrm{rms}}}{\kappa\sigma}\ . 
\end{equation} 
The choices for $Ra$ that we have used imply that
$Re$ is approximately $150$ for each of the four
cases. A summary of our choice of parameters for each case is given in Table~\ref{tab:one}.

All the parameters that have been discussed so far relate to the
hydrodynamic properties of the convection. If all other
non-dimensional parameters in the system are fixed,
the early evolution of any weak magnetic field in the domain depends
solely upon value of $\zeta_0$, which is a parameter that can be set
as the field is introduced. This parameter is proportional to the
magnetic diffusivity, $\eta$, so we require low values of $\zeta_0$
for dynamo action. Equivalently, we could say that dynamo action is only expected in the high magnetic Reynolds number regime. As for the Reynolds number, we shall discuss an alternative definition for the magnetic Reynolds number in Section \ref{sec:dimnum}. However, for the moment, we make an analogous global definition:
\begin{equation}
R_M = \frac{U_{\textrm{rms}}}{\kappa\zeta_0}\ .
\end{equation}
We choose a range of values for $\zeta_0$ which give
values of $R_M$ that vary between approximately $30$ and $480$ ($0.12\le \zeta_0 \le2.0$ for $\theta=3$ and $0.05 \le \zeta_0 \le 0.8$ for $\theta=10$). This range of values is restricted by numerical constraints: smaller values of $\zeta_0$ (higher magnetic Reynolds numbers) would require a higher level of numerical resolution, which would greatly increase the computational expense. 

\subsection{Numerical method}

Solving the equations of compressible convection in the presence of a magnetic field is more demanding numerically than equivalent Boussinesq calculations, so it is important to use a well optimised code. 
The given set of equations is solved using a modified version of the
mixed pseudo-spectral/finite difference code originally described by
\citet{matt95}. Due to periodicity in the horizontal direction,
horizontal derivatives are computed in Fourier space using fast
Fourier transforms. In the vertical direction, a fourth-order finite
differences scheme is used, using an upwind stencil for the advective
terms. The time-stepping is performed by an explicit third-order Adams-Bashforth technique, with a variable time-step. For all
calculations presented here, the aspect ratio is $\lambda=4$. The
resolution is $256$ grid-points in each horizontal directions and
$120$ grid-points in the vertical direction. A poloidal-toroidal
decomposition is used for the magnetic field in order to ensure that
the field remains solenoidal. As explained later, we also aim to
calculate Lagrangian statistics. Trajectories of fluid particles are
therefore computed using the following equation: 
\begin{equation}
\label{eq:part}
\frac{\partial \bm{x}_p}{\partial t}=\bm{u}(\bm{x}_p,t)\ ,
\end{equation}
where $\bm{x}_p$ is the position of the particle.
The velocity at the particle position is interpolated from the grid
values using a sixth-order Lagrangian interpolation scheme. The
boundaries are treated with a decentred scheme and we carefully check
that all the particles remain in the fluid domain. The time-stepping
used to solve Equation~\eqref{eq:part} is the same as for the
other evolution equations in the system.

%
%
\section{Hydrodynamical considerations \label{sec:hydro}}

As we have already described, fully-developed hydrodynamic convection
is used as a starting point for all dynamo calculations. In this
section, we consider the properties of the hydrodynamical flows that
are obtained in each of four cases that are listed in Table~\ref{tab:one}.

\subsection{Dimensionless numbers \label{sec:dimnum}}

In a stratified layer, most quantities of interest will be a function of
depth. This is true not only for the thermodynamic quantities in
the flow, but also for some of the parameters in the system. In the
previous section, we defined the Taylor number and the Reynolds number
using mid-layer values for the density, in addition to using the layer
depth for the characteristic length scale. These are certainly
valid definitions for these quantities, but further insight can be
gained by considering the depth-dependence of these parameters,
particularly when comparing calculations with different levels of
stratification. This is a point that we consider in detail in this
subsection.   

\begin{figure}
\unitlength 0.5mm
\begin{picture}(200,150)
        \put(-4,0){\includegraphics[height=140\unitlength]{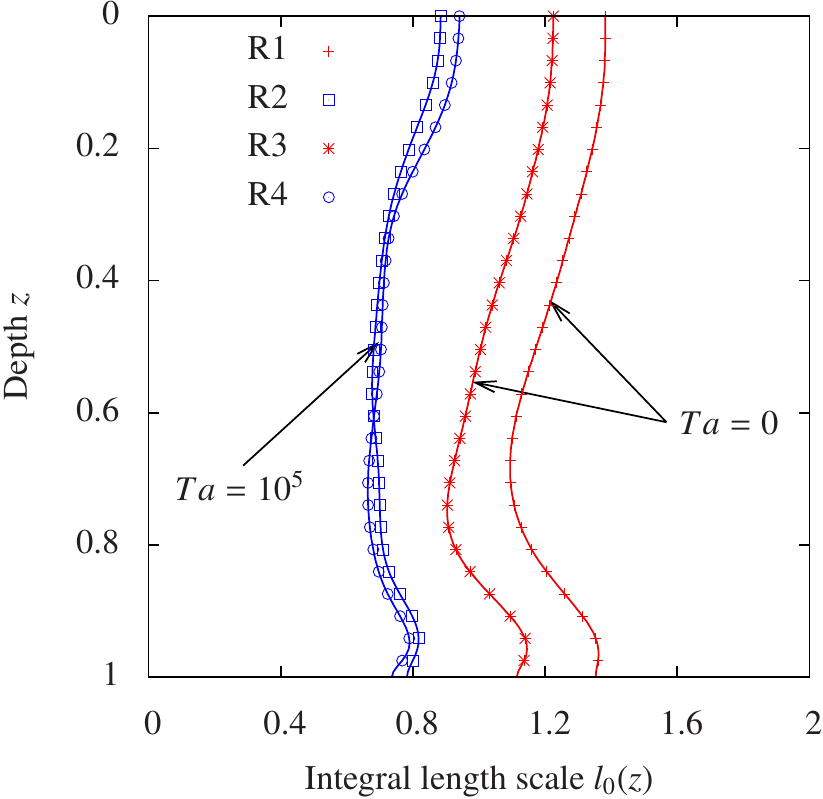}}
        \put(142,0){\includegraphics[height=140\unitlength]{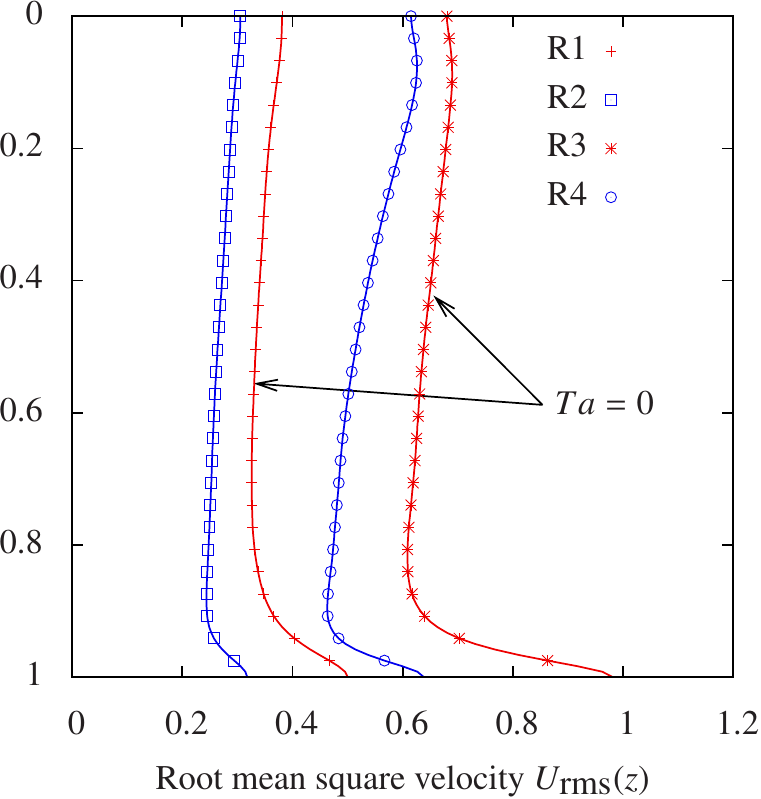}}
	\put(126,130){(a)}
	\put(260,130){(b)}
\end{picture}
\caption[]{Variation of (a) the integral length scale $l_0(z)$ and (b)
  the root mean square velocity $U_{\textrm{rms}}(z)$ with depth for
  the four calculations considered here. The rotating cases correspond
  to the empty blue symbols whereas the non-rotating cases correspond
  to the red crosses.}
\label{fig:l0_urms}
\end{figure}

For each value of $z$, it is possible to define \textit{local}
dimensionless numbers by considering horizontally-averaged
quantities. Under such circumstances, it is more sensible to define
these parameters in terms of a horizontal length-scale rather than
using the depth of the layer (which equals unity in this dimensionless system). We choose here the horizontal integral length scale, defined by 
\begin{equation}
l_0(z)=\frac{1}{\left<u_i(\bm{x},t)u_i(\bm{x},t)\right>_z}\int_0^{\lambda}\left<u_i(\bm{x},t)u_i(\bm{x}+r\bm{e}_x,t)\right>_z\textrm{d}r\ 
\end{equation}
(where we have assumed the the flow is horizontally-isotropic). Here, and in
the following, the brackets $<.>_z$ mean a
  statistical average over horizontal coordinates and time at a given
depth $z$, whereas the brackets $<.>$ (with no
  subscript) mean a statistical average over time and \textit{all} spatial
coordinates. We note that this horizontal scale (unlike the vertical dimension of the domain) will vary
not only with depth, but also from one flow to another.  

Figure~\ref{fig:l0_urms} shows the variation with depth of both the
integral scale, $l_0(z)$, and the local rms velocity,
$U_{\textrm{rms}}(z)=<|\bm{u}(\bm{x},t)|^2>^{1/2}_z$,
for the four cases. Each of these horizontally-averaged quantities is
averaged over more than $100$ acoustic crossing times. In all four
cases, the trends are similar. Near the surface, the integral scale
$l_0(z)$ decreases with depth as the flow becomes more turbulent, and
therefore less correlated. As the flow reaches the lower part of the layer, the
integral length scale increases again, presumably due to the presence
of boundaries. Comparing the two non-rotating cases, one observes that
the effect of increasing $\theta$ is to decrease the integral
scale of the flow. However, whatever the value of $\theta$ is, the
length scales are very similar in the two rotating cases, and are
always smaller than the corresponding integral
scales in the non-rotating calculations. This mirrors
  the result from linear theory
that rotation tends to decrease the size of the convective cells
\citep{chan61}. The rms velocity is comparatively independent of depth,
except close to the lower boundary where (like the integral scale) it increases. Note that
$U_{\textrm{rms}}(z)$ is larger in both the highly-stratified
cases. At fixed stratification, the rms velocity is smaller in the
rotating cases.   

We can now use these velocity and length scales to define local
dimensionless parameters for this system. Unlike their corresponding
global values (based upon the depth of the layer and the depth-averaged rms velocity), these are always functions of $z$. We now define the
\textit{local} Reynolds number to be 
\begin{equation}
Re(z)=\frac{\left<\rho(\bm{x},t)\right>_z U_{\textrm{rms}}(z)l_0(z)}{\kappa\sigma}\ .
\end{equation}
Similarly, we can define the local Rossby number to be 
\begin{equation}
Ro(z)=\frac{U_{\textrm{rms}}(z)}{2\Omega l_0(z)} \ ,
\end{equation}
whilst
\begin{equation}
Ta(z)=\frac{4\left<\rho(\bm{x},t)\right>_z^2\Omega^2l_0^4(z)}{\kappa^2\sigma^2}
\end{equation}
gives a local definition for the Taylor number. Although we are
focusing upon the hydrodynamic properties of convection in this
section, this is also a convenient place to define the local magnetic
Reynolds number, 
\begin{equation}
R_M(z)=\frac{U_{\textrm{rms}}(z)l_0(z)}{\kappa\zeta_0} \ .
\end{equation}
In the dynamo calculations in the next section, we shall often refer to the mid-layer value of the local magnetic Reynolds number, i.e. $R_M(0.5)$.  

\begin{figure}
\unitlength 0.5mm
\begin{picture}(250,140)
        \put(-5,0){\includegraphics[height=130\unitlength]{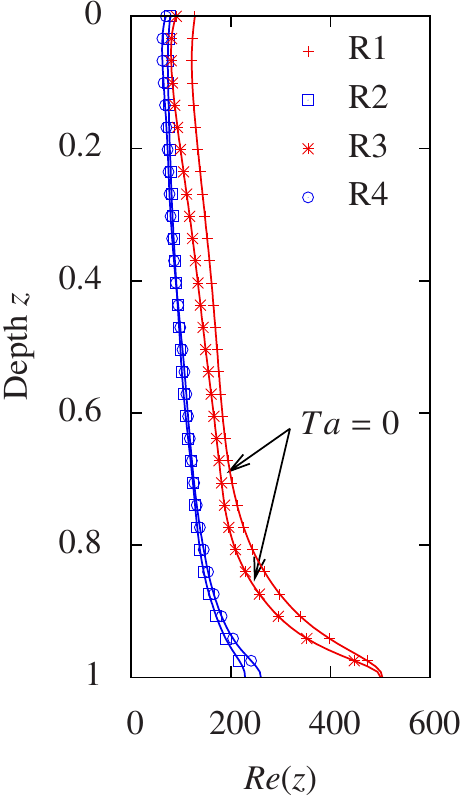}}
        \put(71,0){\includegraphics[height=130\unitlength]{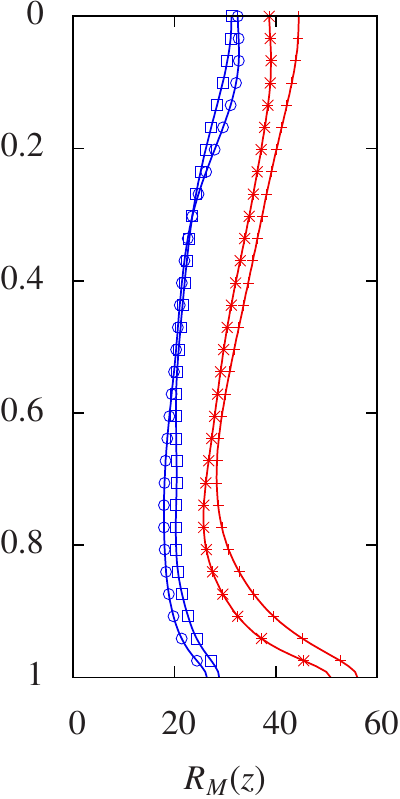}}
        \put(138,0){\includegraphics[height=130\unitlength]{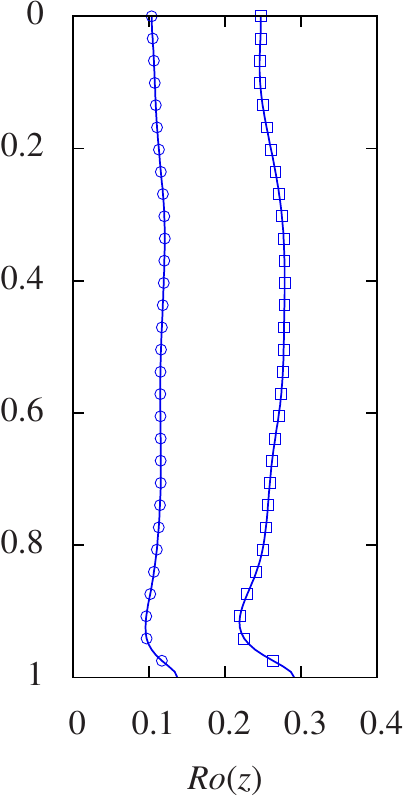}}
        \put(206,0){\includegraphics[height=130\unitlength]{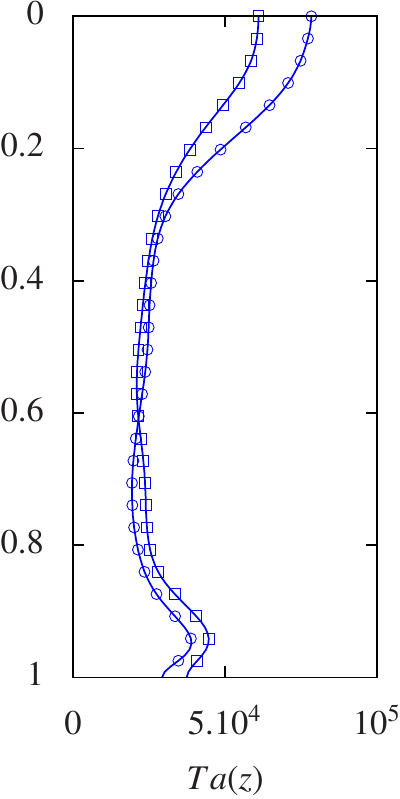}}
\end{picture}
\caption{Dimensionless numbers versus depth. From left to right:
  Reynolds number $Re(z)$, magnetic Reynolds number $R_M(z)$ (the
  global value of $R_M$ is $30$ in each of the four cases shown here), Rossby number $Ro(z)$ and Taylor number $Ta(z)$.\label{fig:dimensionless}}
\end{figure}

The depth-dependence of the local Reynolds number, the local magnetic
Reynolds number (for four cases corresponding to a global value of $R_M=30$), the local Rossby number and the local Taylor number is shown in Figure~\ref{fig:dimensionless}. First of all, it is clear that the local Reynolds number is an increasing function of depth, which is a clear indication that the flow is becoming more turbulent
as $z$ increases.
This variation with depth is partly due to the fact that the dynamic viscosity $\mu$ is held constant in all our simulations.
An important observation is that the local Reynolds number
(and the local magnetic Reynolds number)
is always smaller in the rotating cases. This is
  despite the fact that the \textit{global} Reynolds numbers and
  magnetic Reynolds numbers (as defined in the previous subsection)
  are the same in all cases. This reduction in the local Reynolds
number is due to the fact that rotation leads to motions with a
smaller characteristic horizontal length scale (as well as a lower rms
velocity). So, if these local definitions for the Reynolds numbers
give a clearer indication of the level of turbulence in the flow, the
rotating calculations are actually slightly less turbulent than suggested by the values of the global Reynolds numbers. We shall
return to this point in Section~\ref{sec:dynamo}. Returning to
Figure~\ref{fig:dimensionless}, we see that the Rossby number
is roughly constant across the layer, but is smaller in the high
$\theta$ case. Although we could conclude that this implies that the
effects of the Coriolis force with respect to inertia are more important in the highly stratified
case, the key point is that $Ro(z)$ is always much less that unity,
which implies that the flows are both rotationally dominated. Finally,
the local Taylor number is largely independent of depth (and $\theta$)
in the middle of the domain, but increases near the boundaries. Note
that this behaviour is very different from the global Taylor number,
which actually \textit{increases} with depth. This apparent
discrepancy is due to the fact that the local Taylor number is
strongly influenced by variations in the integral scale (varying as
$l_0^4(z)$), which tend to outweigh any local increases in the mean
density. The Taylor number (in particular) highlights some of the
difficulties that must be faced when defining appropriate dimensionless
numbers in highly-stratified fluids.  

\subsection{Lyapunov exponents \label{sec:lyap}}

From the point of view of dynamo action, it is interesting to quantify
the amount of stretching in the flow. This can be
  achieved by measuring the corresponding Lyapunov exponents. Several
  previous studies have explored the relationship between dynamo
  action and Lyapunov exponents \citep[see, for
  example,][]{finn1988,brand1995,brummell1998b,tanner2003}, and we
  would expect regions with large Lyapunov exponents to be playing a
  dominant role in the dynamo process. To estimate the short-time
  Lyapunov exponent, we release $5\times10^4$ fluid particles
randomly within the convective layer. For each particle, we also
release a second particle at a distance $d_0=10^{-4}$ apart from the
first particle, in a random direction. This pair of particles is then
followed as it moves through the fluid. The short-time Lyapunov
exponent $\lambda_e$ \citep{eckhardt93} is then calculated using the
following expression: 
\begin{equation}
\label{eq:le}
\lambda_e=\frac1t\log \frac{d(t)}{d_0} \ ,
\end{equation}
where $d(t)$ is the distance between the two particles at time $t$. The Lyapunov exponent is computed after a fixed time (approximately equal to one crossing time) and the distance between the particles is then renormalised to $d_0$.
The results appear to be insensitive to the choice of the initial
separation distance $d_0$. Furthermore, they do not appear to depend
upon the details of the method that is used to reinitialise the
particle positions. All initial separations will quickly align with
the expanding manifold, so we would expect to see very little
dependence upon the initial condition for sufficiently large time,
$t$. 
\begin{figure}
\unitlength 0.5mm
\begin{picture}(250,220)
        \put(-5,112){\includegraphics[height=100\unitlength]{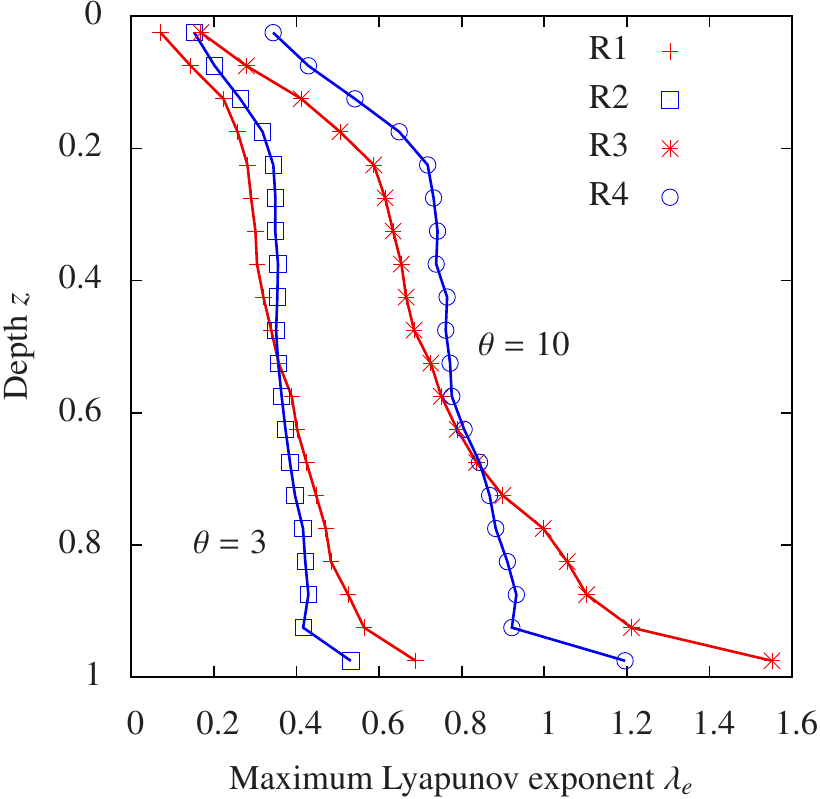}}
        \put(-5,3){\includegraphics[height=100\unitlength]{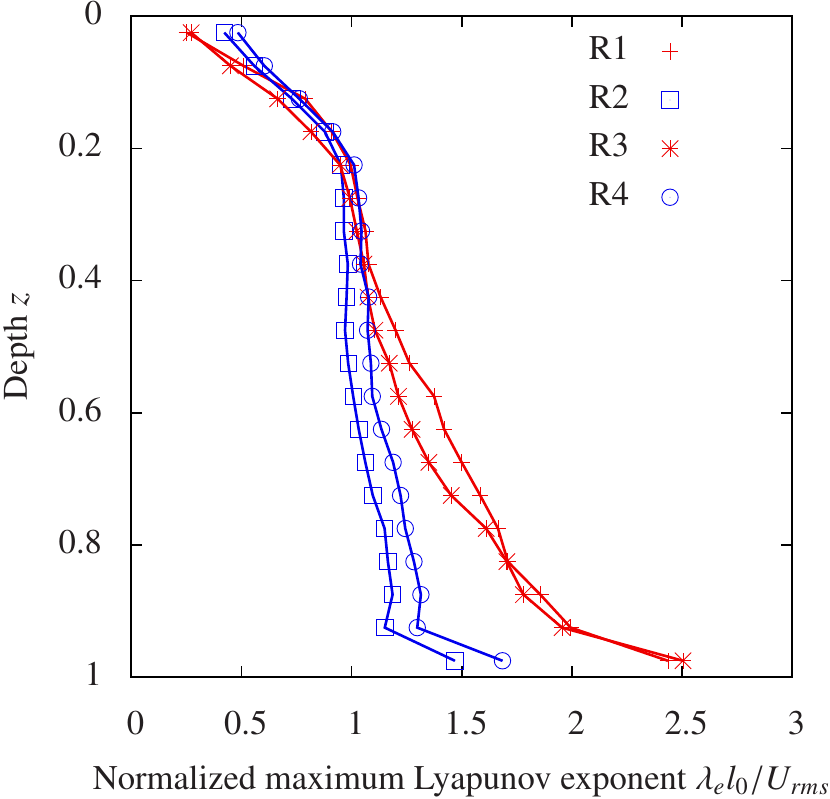}}
        \put(110,166){\includegraphics[height=48\unitlength]{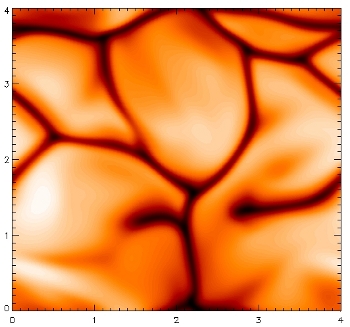}}
        \put(163,166){\includegraphics[height=48\unitlength]{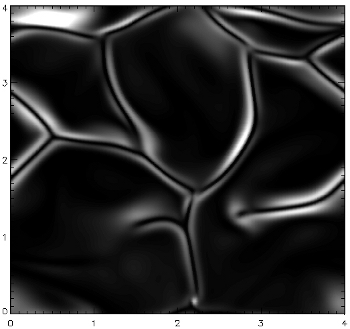}}
        \put(216,166){\includegraphics[height=48\unitlength]{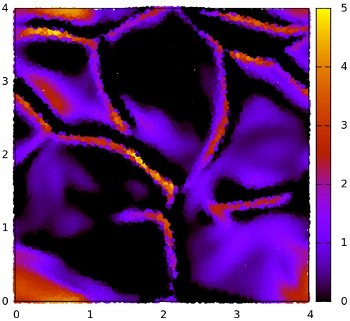}}
        \thicklines
        \put(110,116){\includegraphics[height=48\unitlength]{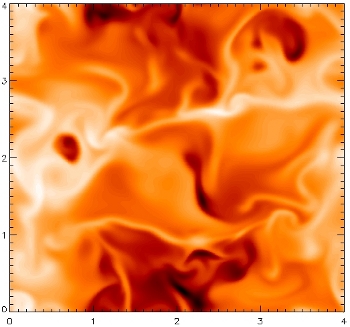}}
        \put(163,116){\includegraphics[height=48\unitlength]{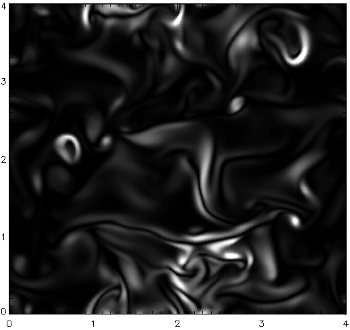}}
        \put(216,116){\includegraphics[height=48\unitlength]{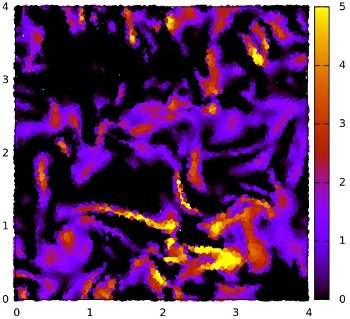}}
%
        \put(110,66){\includegraphics[height=48\unitlength]{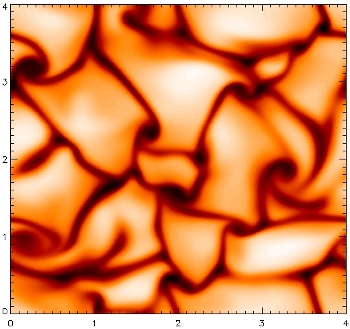}}
        \put(163,66){\includegraphics[height=48\unitlength]{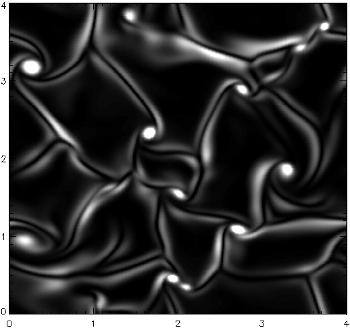}}
        \put(216,66){\includegraphics[height=48\unitlength]{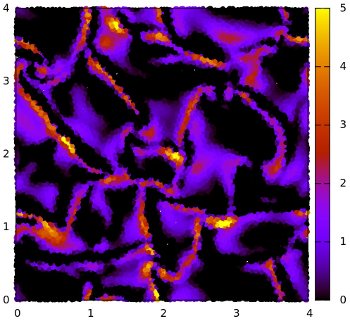}}
        \put(110,16){\includegraphics[height=48\unitlength]{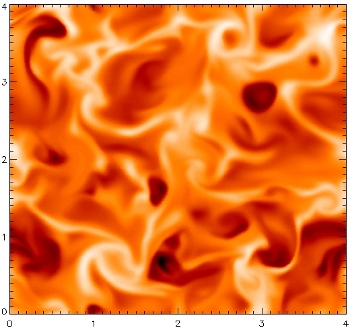}}
        \put(163,16){\includegraphics[height=48\unitlength]{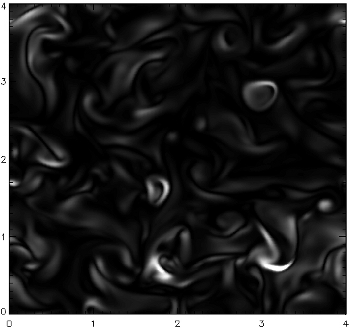}}
        \put(216,16){\includegraphics[height=48\unitlength]{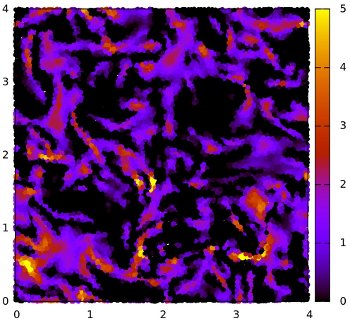}}
%
        \put(118,6){\small{Temperature}}
        \put(177,6){\small{Enstrophy}}
        \put(233,6){\small{STLE}}
	\put(83,202){(a)}
	\put(83,93){(b)}
%
%
\end{picture}
\caption{(a) Maximum Lyapunov exponents, $\lambda_e$, versus depth. (b) Maximum Lyapunov exponents scaled by the mean turn-over time $l_0/U_{\textrm{rms}}$. Right column: Comparison between non-rotating (two first rows) and rotating convection (two last rows). From left to right: Temperature fluctuations, enstrophy and short time Lyapunov exponents (STLE)  at a given depth $z$. From top to bottom: $z=0.2$ without rotation, $z=0.8$ without rotation, $z=0.2$ with rotation, $z=0.8$ with rotation. $\theta=3$ in all cases.\label{fig:lyapplot}}
\end{figure}
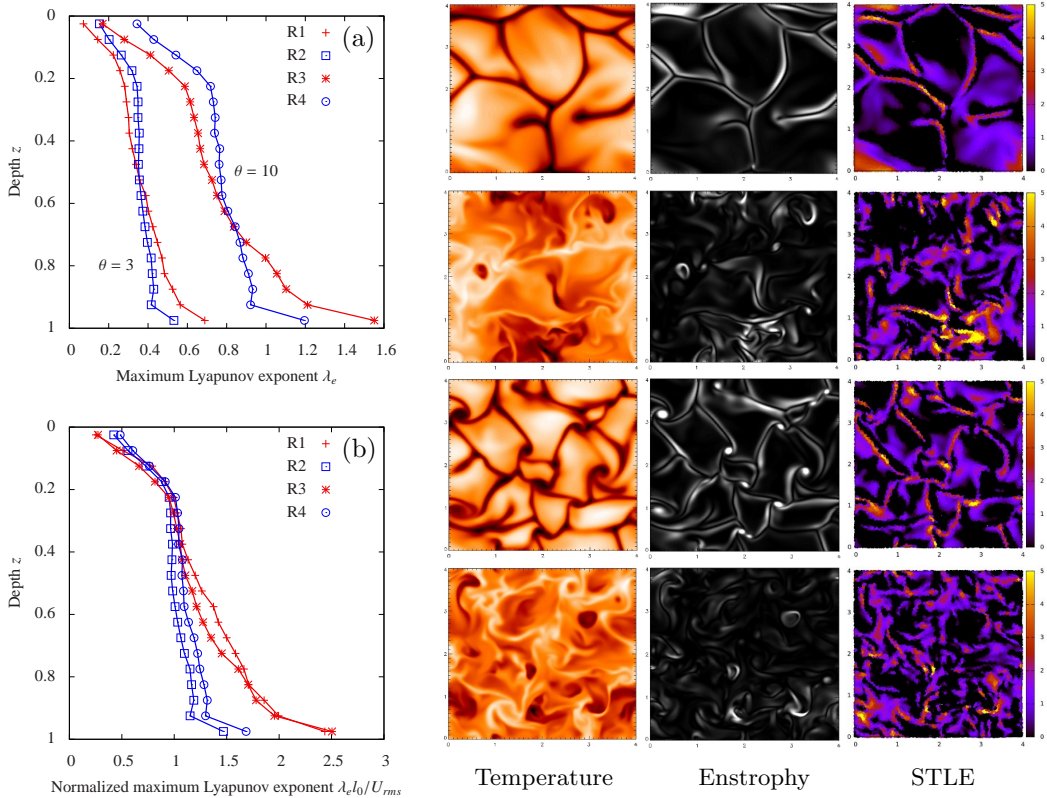

To study the depth dependence of the stretching within the flow, we consider the horizontal average of the largest Lyapunov exponent at each value of $z$.
In addition, we time-average the resulting Lyapunov exponent over approximately $20$ convective turnover times.
Figure~\ref{fig:lyapplot}(a) shows the mean maximum Lyapunov exponent versus
depth for each of the four cases. In all cases, the stretching
increases with depth (as the flow becomes more and more turbulent),
with the Lyapunov exponents taking their maximum values at the bottom
of the convective domain. Unsurprisingly, this is reminiscent of the
depth-dependence of the local Reynolds number. This depth-dependence has
important implications for the dynamo problem: we would expect the
magnetic field to be mostly generated in the lower part of the layer,
where the rate of stretching is maximal. Compared to the
equivalent non-rotating cases, in rotating convection we see higher
values of the Lyapunov exponent at the top of the layer and lower
values at the bottom of the layer. In other words, the maximum Lyapunov
exponent is less
depth-dependent in the rotating cases than it is in the non-rotating
calculations. From a dynamo perspective, this suggests that we should expect the magnetic fields
to be more homogeneously distributed across the layer in the rotating
cases than they are in the non-rotating simulations. Figure~\ref{fig:lyapplot}(b)
shows a plot in which the Lyapunov exponents are scaled by
the local convective turnover time  $l_0(z)/U_{\textrm{rms}}(z)$. We
use a similar scaling for the dynamo growth rates in the next section,
where the implications of this plot will be discussed in more
  detail. Here we simply note two key features of this scaled
plot. Firstly, this scaling causes the four curves to collapse onto a
single curve in the upper part of the layer. Secondly, we note that
the scaled Lyapunov exponents suggest that there is less stretching in
the lower part of the domain in the rotating cases.

On the right-hand side of Figure~\ref{fig:lyapplot}, we show a snapshot of
the temperature, enstrophy and the short time Lyapunov exponent (STLE)
in a horizontal plane at two different depths ($z=0.2$ and $z=0.8$) for
both non-rotating (top) and rotating (bottom) convection in the
weakly-stratified ($\theta=3$) case. In the temperature plot, brighter
contours correspond to warmer regions of fluid. As expected from our considerations of the horizontal integral scale,
it is clear that the horizontal scale of convection is smaller in the
rotating case than it is in non-rotating convection. This reduction in horizontal scale is also apparent in the
enstrophy plot where bright regions, corresponding to areas of high
(squared) vorticity, outline the edges of the
  convective cells. The STLE map is obtained by releasing $10^5$
particle pairs at a given depth. The particles are followed for
approximately one crossing time. The STLE is then computed using
Equation~\eqref{eq:le} and plotted at the initial particle pair
position. Comparing the STLE map with the enstrophy
  distribution, it is clear that there is a correlation between
zones of strong stretching and regions of high
vorticity. Given this correlation, it is natural to conclude
that the higher (unscaled) Lyapunov exponent in the upper part of the
rotating simulations, as observed in Figure~\ref{fig:lyapplot}(a), is
mostly due to the larger number of convective cells that are
present. The scaled plot that is shown in
  Figure~\ref{fig:lyapplot}(b) tends to support this conclusion: the
scaling that has been used here takes into account this ``filling
factor'' effect, which explains why all the scaled Lyapunov exponent
curves collapse onto a single curve near the surface in that case.  

%
%

\section{Dynamo simulations\label{sec:dynamo}}

In this section, we investigate the dynamo properties
  of the four convective flows that are being considered in this
  paper. Each dynamo calculation is initialised by introducing a seed
  magnetic field into statistically-steady hydrodynamic
  convection. The initial spatial structure of this magnetic field is
  given by $\bm{B}(x,y,z)=$($A\cos(k_i y)$, $A\cos(k_ix)$, $0$), where
  $k_i=2\pi/\lambda$ and $A$ is the peak amplitude of the initial
  perturbation. This is almost the simplest possible magnetic field
  configuration that is consistent with the imposed boundary
  conditions, whilst also ensuring that there is no net (horizontal or
  vertical) magnetic flux across the computational domain. Test
  calculations suggest that the evolution of the dynamo is largely
  insensitive to the precise choice of initial conditions.

\subsection{The kinematic regime\label{sec:kinematic}}

Initially, we consider the kinematic dynamo regime
in which the seed magnetic field is assumed to be weak. This implies
that the magnetic field tends to grow (or decay) exponentially at a
rate that is determined by the value of magnetic Reynolds
number. Given that the actual growth of the
magnetic energy fluctuates in time, long
time-averages are often needed in order to accurately measure growth
rates. This kinematic phase of evolution can be indefinitely prolonged by
switching off the Lorentz force terms in the momentum equation and the
ohmic heating terms in the temperature equation. This is the procedure
that is adopted in this subsection. For each
velocity field, we carry out $6$ kinematic
  calculations at different values of $\zeta_0$. This parameter is
  chosen so that the global magnetic Reynolds number ranges from
  approximately $30$ to $480$ in each case. This implies that the
magnetic Prandtl number (the ratio of the global magnetic Reynolds
number to the global Reynolds number) varies from approximately $0.2$
to $3.2$.   
\begin{figure}
\unitlength 0.5mm
\begin{picture}(250,260)
        \put(2,-3){\includegraphics[height=118\unitlength]{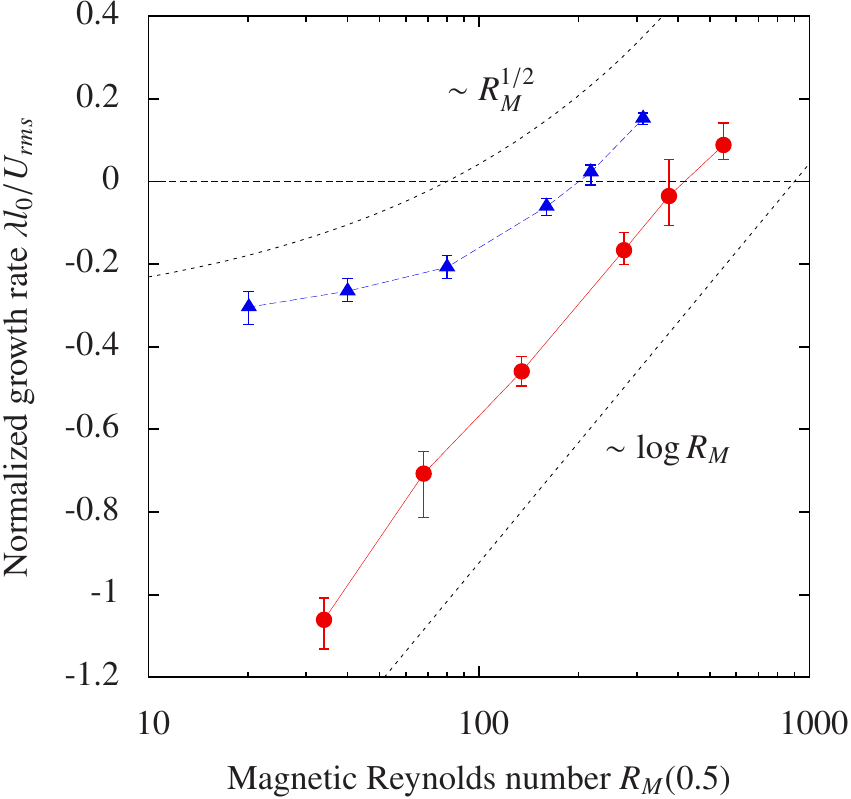}}
        \put(148,-3){\includegraphics[height=118\unitlength]{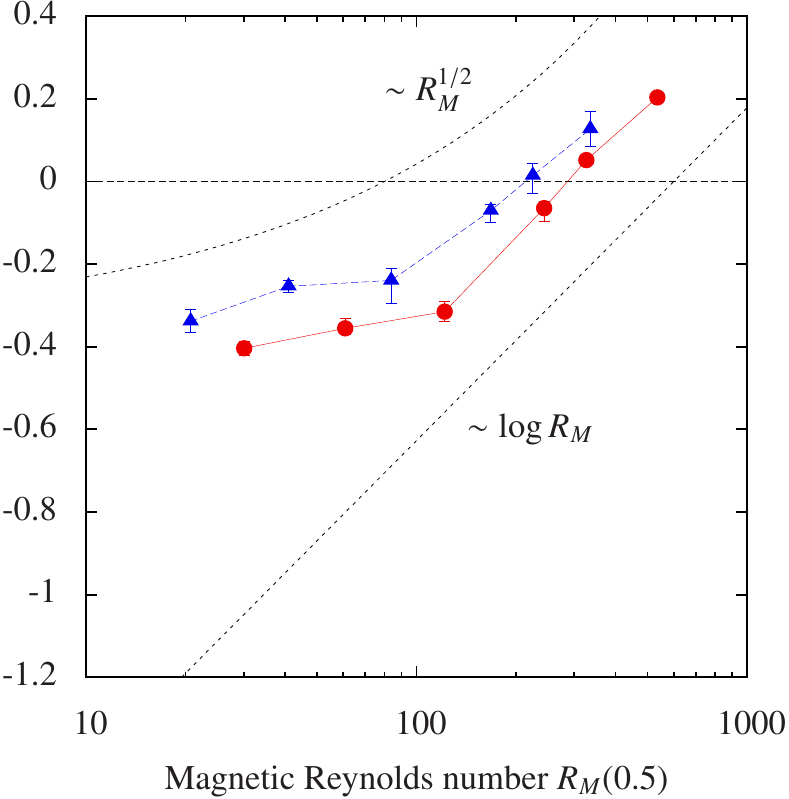}}
        \put(3,120){\includegraphics[height=130\unitlength]{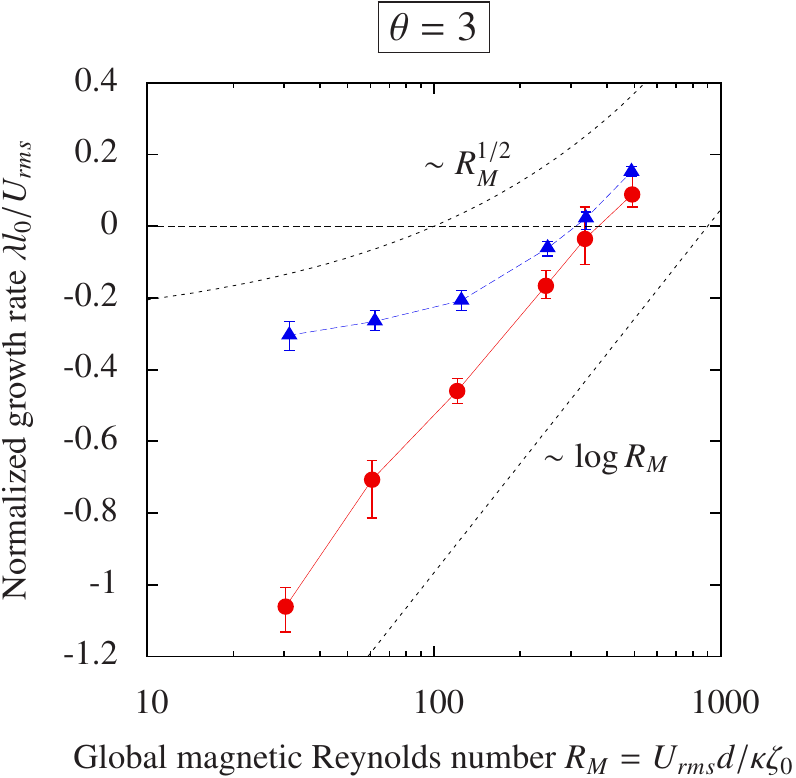}}
        \put(149,120){\includegraphics[height=130\unitlength]{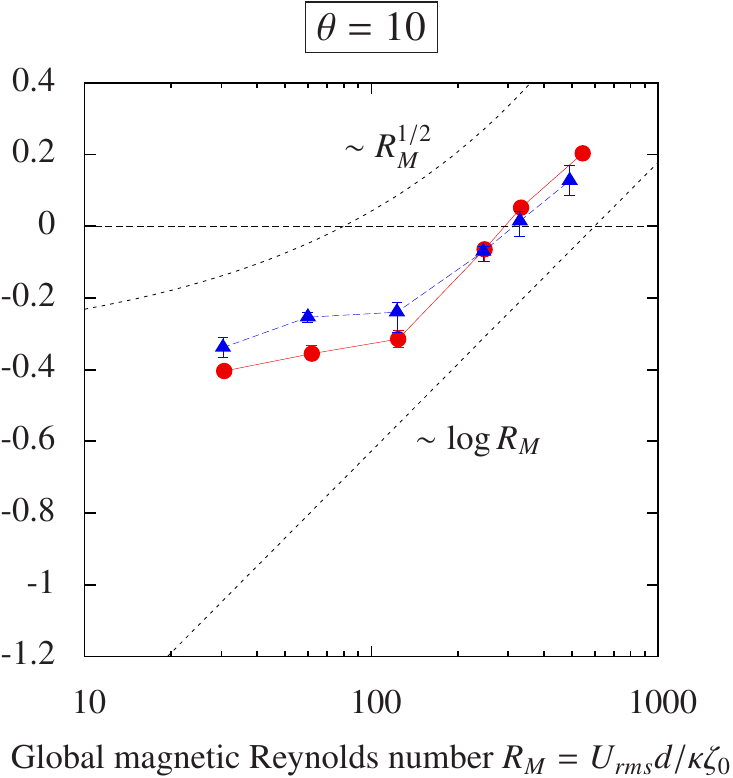}}
\end{picture}
\caption{Growth rates of the magnetic energy versus the magnetic Reynolds number for $\theta=3$ (left) and $\theta=10$ (right). In the upper plots, the global magnetic Reynolds number has been used. In the lower plots, the same growth rates have been plotted against the mid-layer value of the local magnetic Reynolds number, $R_M(0.5)$. The rotating results are plotted with triangles whereas the non-rotating results correspond to the circles. The growth rates are normalised by the mid-layer turnover-time $l_0(0.5)/U_{\textrm{rms}}(0.5)$. The error bars correspond to the maximum and minimum value for the growth rate when considering different time intervals. The logarithmic and square-root scalings are plotted for reference.\label{fig:growth}}
\end{figure}

The main aim of these kinematic calculations is to determine the ways
in which rotation and stratification influence the rate of growth (or
decay) of the seed magnetic field. The upper part of
Figure~\ref{fig:growth} shows two plots (for $\theta=3$ and
$\theta=10$) of the kinematic growth rate of the magnetic energy,
$\lambda$, versus the global magnetic Reynolds number. In all cases,
$\lambda$ has been scaled by the mid-layer turnover time of the turbulence,
$l_0(0.5)/U_{\textrm{rms}}(0.5)$. This rescaling has been carried out to
facilitate comparison between the four cases, as well
  as the Lyapunov exponent calculations from the previous
  section. Results from this kinematic study highlight several key
points. Firstly, it is clear that the critical value (for the onset of
dynamo action) of the global magnetic Reynolds number is rather
similar in all four cases. It is only in the
  non-rotating $\theta=3$ case that there is a suggestion of a
  slightly higher critical global magnetic Reynolds number (although
  the error bars in the growth rate close to criticality are large enough to suggest that this difference may not be significant). Adopting this definition for the magnetic Reynolds number, we see that the scaled growth rates in the highly stratified case seem to be rather insensitive to
the presence of rotation. It is only in the weakly stratified case, at
small values of the global magnetic Reynolds number, that significant
differences are seen between the rotating and the non-rotating
cases. So it is tempting to conclude from these results
  that the dynamo properties of compressible convection (particularly
  at higher values of the global magnetic Reynolds number) are largely
  insensitive to the presence of rotation, as well as variations in
  the level of stratification.

However, some caution is needed when interpreting
  these results. As we discussed in the previous section, a
  global definition for the magnetic Reynolds number takes no account of the
horizontal scale of convection (a quantity that varies significantly
between the four cases that are being considered). So we would 
argue that it is more sensible to consider representative values of
the \textit{local} magnetic Reynolds number for the purposes
  of this comparison. The lower part of Figure~\ref{fig:growth} also
shows the growth rate curves for the four different cases, but this time
$\lambda$ has been plotted against the mid-layer value of the local
magnetic Reynolds number, $R_M(0.5)$. It is now
  possible to discern clear differences between the four sets of
  calculations. We first discuss the effect of varying $\theta$
without rotation. For $\theta=3$, the critical value for $R_M(0.5)$ is
about $420$, whereas for $\theta=10$, the critical value for
$R_M(0.5)$ is about $290$. Therefore, an increase in the
level of stratification reduces the critical value of the local magnetic
Reynolds number. Let us now consider the two rotating cases. A
striking feature of the plots in the lower part of
Figure~\ref{fig:growth} is how similar the growth rate curves of the two
rotating cases are. For both values of $\theta$, the critical value of
$R_M(0.5)$ is about $220$. Therefore, independently of
the level of stratification, rotation tends to reduce the critical
value for the local magnetic Reynolds number. 

Whatever definition is adopted for the magnetic
  Reynolds number, it is clear that the growth rate curves exhibit
  certain characteristic features. The dotted lines on
Figure~\ref{fig:growth}, which are shown on these plots for
indication, correspond to a logarithmic scaling \citep[see, for
example,][]{roga97} and a square-root scaling \citep{scheko04} of the
growth rate with the magnetic Reynolds number. A logarithmic scaling
seems to be valid (at least for all magnetic Reynolds numbers
considered here) for the non-rotating $\theta=3$ case. This situation
is less clear in the highly-stratified non-rotating case, although an
$R_M^{1/2}$ scaling may be more appropriate here. The other
possibility is that there are two different regimes in this case, with
a logarithmic scaling holding only for larger values of the magnetic Reynolds
number. The scalings are again not completely
  convincing in the rotating cases, but the data seem to be compatible with an
  $R_M^{1/2}$ scaling, irrespective of the value of $\theta$. In both
  plots in the lower part of Figure~\ref{fig:growth} it is clear that
the rotating cases always have higher growth rates than the equivalent
non-rotating calculations. The largest difference in growth rates
between the rotating and non-rotating calculations can
be seen at low magnetic Reynolds number in the $\theta=3$ case. 

\begin{figure}
\unitlength 0.5mm
\begin{picture}(250,140)
        \put(-5,0){\includegraphics[height=130\unitlength]{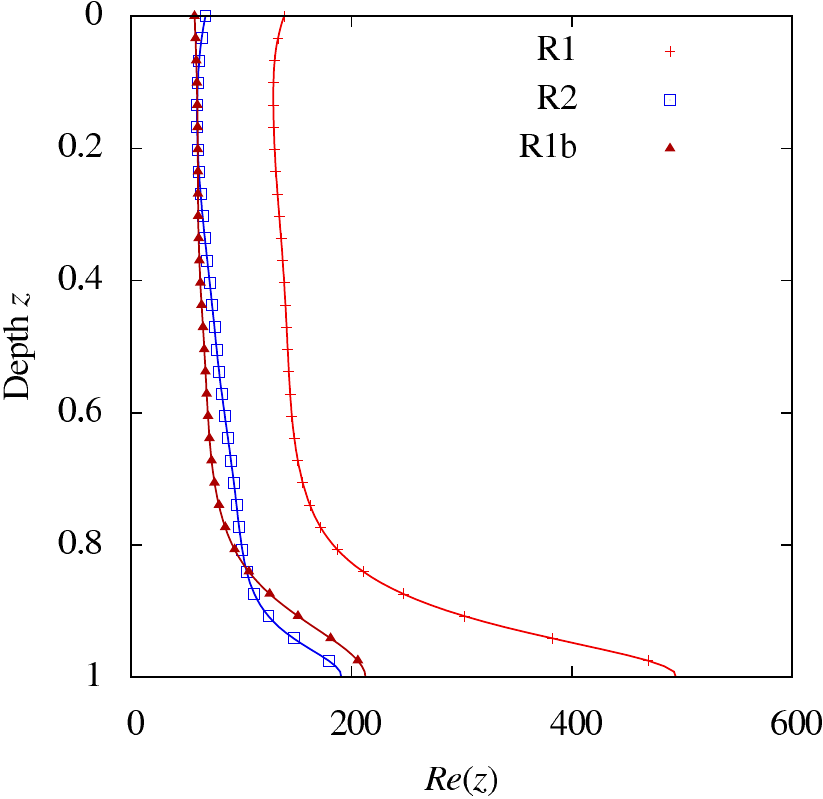}}
        \put(140,0){\includegraphics[height=130\unitlength]{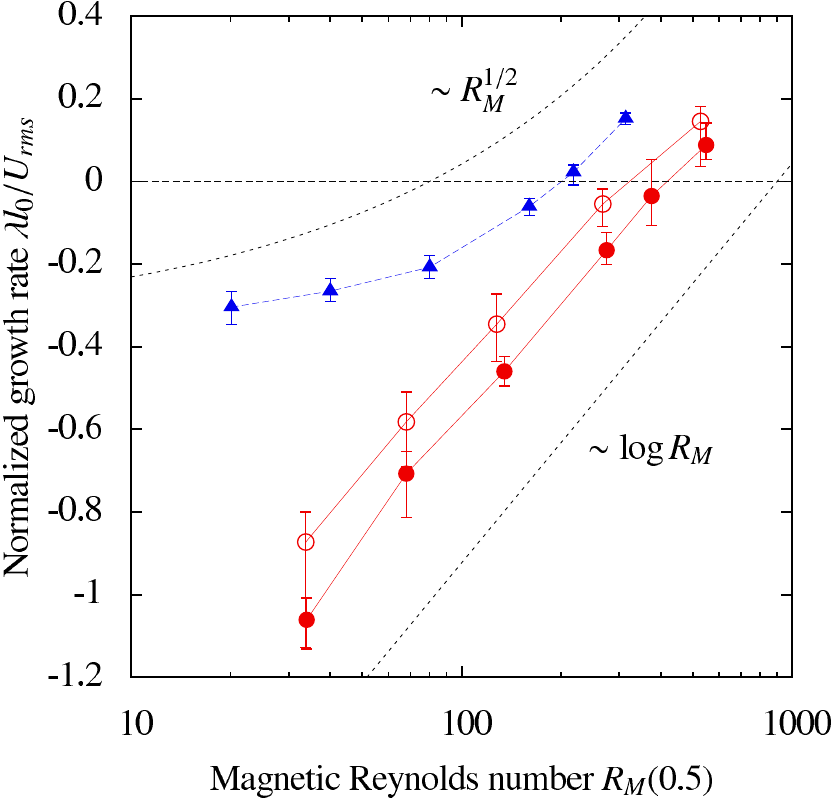}}
	\put(114,26){(a)}
	\put(251,26){(b)}
\end{picture}
\caption[]{Results from an additional set of
    non-rotating simulations at lower local Reynolds number. (a) The
    local Reynolds numbers versus depth for simulations $R1$, $R1b$
    and $R2$. (b) Growth rates of the magnetic energy versus the
    mid-layer value of the local magnetic Reynolds number, $R_M(0.5)$,
    for all the $\theta=3$ simulations. The rotating results, from
    simulation $R2$, are plotted with triangles whereas the
    non-rotating results correspond to the circles (filled for $R1$,
    and empty for $R2$). In all cases, the growth rates are normalised
    by the mid-layer turnover-time
    $l_0(0.5)/U_{\textrm{rms}}(0.5)$.\label{fig:lowra}} 
\end{figure}

The observed variations in the
growth rate curves suggest that rotation is beneficial for dynamo
  action. However, as discussed in the previous section, it could be
  argued that the rotating calculations are (in some sense) less
  turbulent that their non-rotating counterparts: although the global Reynolds
numbers are similar in all cases, the \textit{local} Reynolds number is
significantly smaller in the rotating calculations. So we have to
consider the possibility that it is actually the difference in the
local Reynolds numbers that is giving the impression of enhanced dynamo action in the rotating cases.
In order to address this issue, we performed
an additional set of kinematic simulations for a new convective flow
(referred to here as case $R1b$). Setting $Ta=0$, we choose $\theta=3$
since this it is in this case that we observe the greatest difference
between the dynamo properties of rotating and non-rotating
convection. All parameters are identical to case $R1$ (as given in
Table~\ref{tab:one}), except the Rayleigh number, which we reduce to
$Ra=5\times10^4$. The global Reynolds number is now significantly smaller than in all other calculations reported in this paper. However,
the local value now has a similar depth dependence to the equivalent
rotating case, $R2$, as illustrated in
Figure~\ref{fig:lowra}(a). Figure~\ref{fig:lowra}(b) shows a
comparison of the kinematic dynamo growth rates for all the $\theta=3$
calculations ($R1$, $R1b$ and $R2$). At similar local magnetic
Reynolds numbers, the growth rates for the lower Rayleigh number case
are slightly higher than the equivalent non-rotating higher Rayleigh
number case ($R1$). In other words, the reduction in the local Reynolds number has increased the efficiency of the dynamo. However, there is still a significant discrepancy
between these growth rates and those found in the rotating calculations. This
indicates that the reduction in the critical value of $R_M(0.5)$ in
the rotating cases cannot be explained simply by the differences between the
local Reynolds numbers. Therefore this change in the critical local magnetic Reynolds number must be due to the effects of rotation.

Before concluding this section on kinematic growth
  rates, it is worth commenting on the magnitudes of the growth rates in the dynamo regime. Obviously the
  range of magnetic Reynolds numbers is restricted by numerical
  constraints. However, it is worth noting that our positive growth
  rates (when properly normalised) are comparable to the values
  reported by \citet{kapyla2008}, who also considered simulations of 
  compressible convection (albeit with overshoot and shear), and those
  found in the forced homogeneous turbulence calculations of
  \citet{haugen2004}. It is also of interest to compare the peak
  kinematic growth rates to the scaled Lyapunov exponents that are
  shown in Figure~\ref{fig:lyapplot}(b). In all cases, it is clear that the
  growth rates are significantly smaller than the corresponding
  Lyapunov exponents. This is unsurprising given that the magnetic
  Reynolds numbers are comparatively modest in these numerical
  simulations, and we would expect to see higher growth rates at
  higher values of $R_M$. Nevertheless our results would appear to be
  consistent with the findings of \citet{brand1995} who, for a related
  dynamo calculation, found that the Lyapunov exponents were
  systematically larger than the observed kinematic growth rates.

%
%
\subsection{Magnetic fields in the nonlinear regime\label{sec:nonlinear}}
\begin{figure}
\unitlength 0.5mm
\begin{picture}(250,150)
        \put(0,0){\includegraphics[height=140\unitlength]{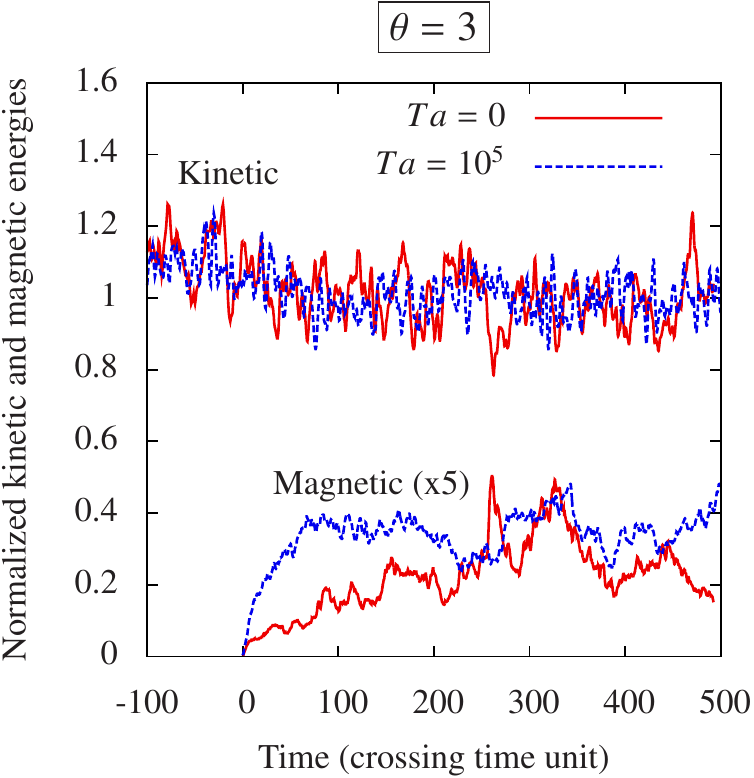}}
        \put(145,0){\includegraphics[height=140\unitlength]{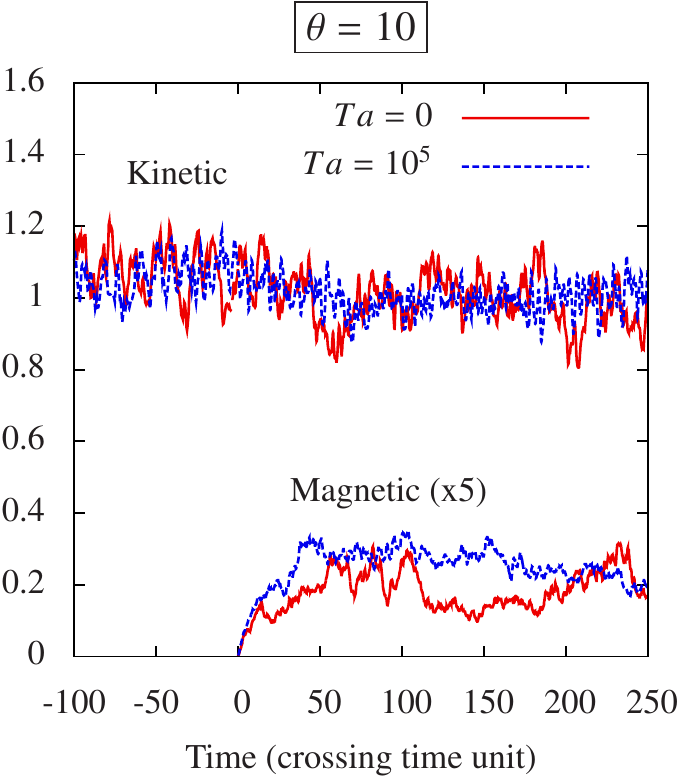}}
	\put(118,26){(a)}
	\put(251,26){(b)}
\end{picture}
\caption[]{The time evolution of the normalised mean kinetic and mean
  magnetic energy densities for (a) $\theta=3$ and (b) $\theta=10$. Both
  quantities are normalised by the mean kinetic energy density during the
  saturated phase (\textit{i.e.} from $t=200$ to $t=500$ in the
  $\theta=3$ cases, and from $t=100$ to $t=250$ in the $\theta=10$
  cases). The magnetic energy has been multiplied by 5.\label{fig:magnl}}
\end{figure}

We now consider the fully nonlinear set of governing equations
(Equations~\eqref{eq:mass} to \eqref{eq:heateq}) including the
back-reaction of the magnetic field on the velocity field. For each of
the four cases under consideration, we choose the values of $\zeta_0$
corresponding to the largest values of the magnetic Reynolds number,
in order to maximise the growth rate of the dynamo, thus minimising
the duration of the kinematic phase. In the following, the time $t=0$
corresponds to the insertion time of the small magnetic
perturbation. The initial ratio between the total magnetic energy in
the seed field and the total kinetic energy in the flow is roughly
$10^{-3}$ in all cases. This implies that the initial field is weak
enough to be kinematic, without unnecessarily extending the kinematic
phase in these nonlinear calculations. Each nonlinear
  calculation has been evolved over a significant fraction of the
  ohmic decay time (based upon the magnetic diffusivity and the
  horizontal integral scale).

\par Figure \ref{fig:magnl} shows the evolution of the mean kinetic and
mean magnetic energy densities for the four cases. All of these
dynamos are highly time-dependent, but there are clear patterns of
behaviour. During the early stages of evolution, the magnetic
perturbation grows. However, the kinematic phase of the dynamo is
extremely brief in these calculations, as the seed field rapidly
becomes dynamically significant. After the short kinematic phase, the dynamo
undergoes a more extended period of nonlinear growth, eventually
settling down to a time-dependent saturated state. Conservation of energy implies that the kinetic energy decreases
during the nonlinear phase of the dynamo, reaching a final level that
is clearly lower than the kinetic energy of the initial hydrodynamic
state. Across all simulations, the mean magnetic energy density,
$\left<|\bm{B}|^2\right>/2$, saturates at a level that is somewhere
between 4 and 9\% of the mean kinetic energy density,
$\left<\rho|\bm{u}|^2\right>/2$. Although not shown
  here, a corresponding nonlinear calculation for the $R1b$ case
  (described in the previous section) saturates at a similar level.
These nonlinear results suggest that the saturation level of the
dynamo is comparatively insensitive to the level of stratification
within the domain. There is weak evidence to suggest that the rotating
cases are saturating at a slightly higher level (particularly for
$\theta=3$), despite the fact that the mid-layer values of the local
magnetic Reynolds number are actually smaller in the rotating
calculations than they are in the corresponding non-rotating
cases. If we were comparing nonlinear calculations at
  similar values of the (mid-layer) local magnetic  Reynolds number,
  as opposed to similar values of the global $R_M$, we would expect the
  rotating cases to saturate at a higher level. However, we did not
  investigate this parameter regime here given that this would require
  higher spatial resolution in the rotating cases.

Looking again at Figure~\ref{fig:magnl}, it is worth noting that the
time-dependence of the magnetic energy in the non-rotating $\theta=3$
case is strongly intermittent. This is clearly seen at $t\approx270$, where a burst of magnetic energy corresponds to a
drop in the kinetic energy. These fluctuations in the magnetic energy
make it more difficult to determine the level at which the dynamo
saturates in this case. As discussed in the previous
  subsection, this non-rotating $\theta=3$ case may have a
  \textit{slightly} higher critical global magnetic Reynolds number
  than the other three cases. If the dynamo is indeed closer to
  criticality than the other cases, we would expect it to be more
  sensitive to time-dependent variations in the flow. This would explain
  the observed behaviour. 

\begin{figure}
\unitlength 0.5mm
\begin{picture}(250,260)
        \put(0,128){\includegraphics[height=130\unitlength]{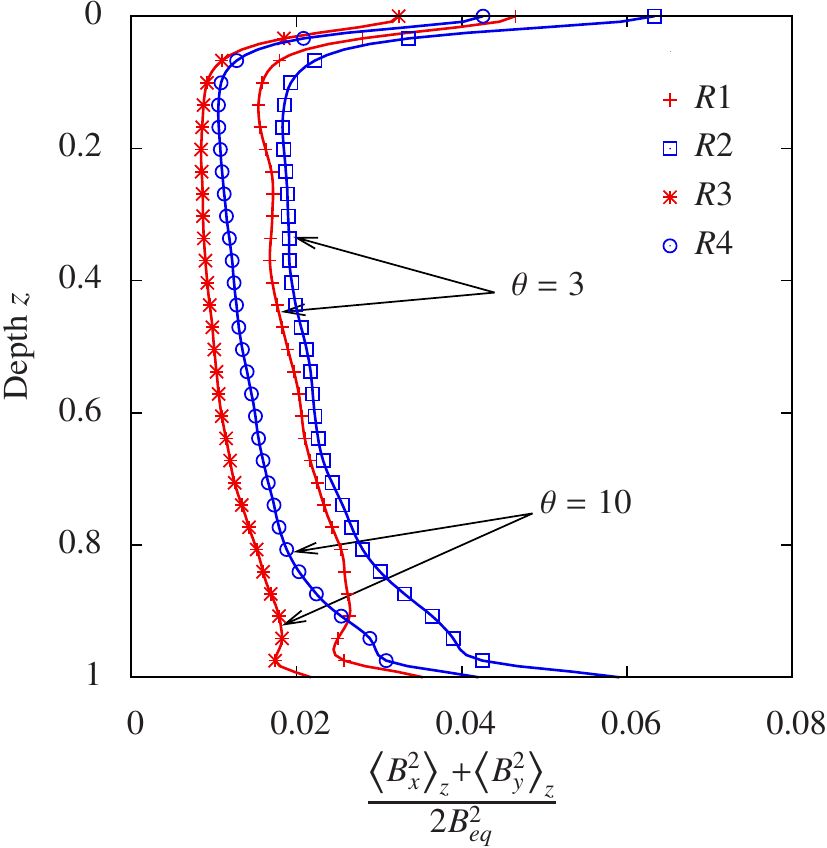}}
        \put(145,128){\includegraphics[height=130\unitlength]{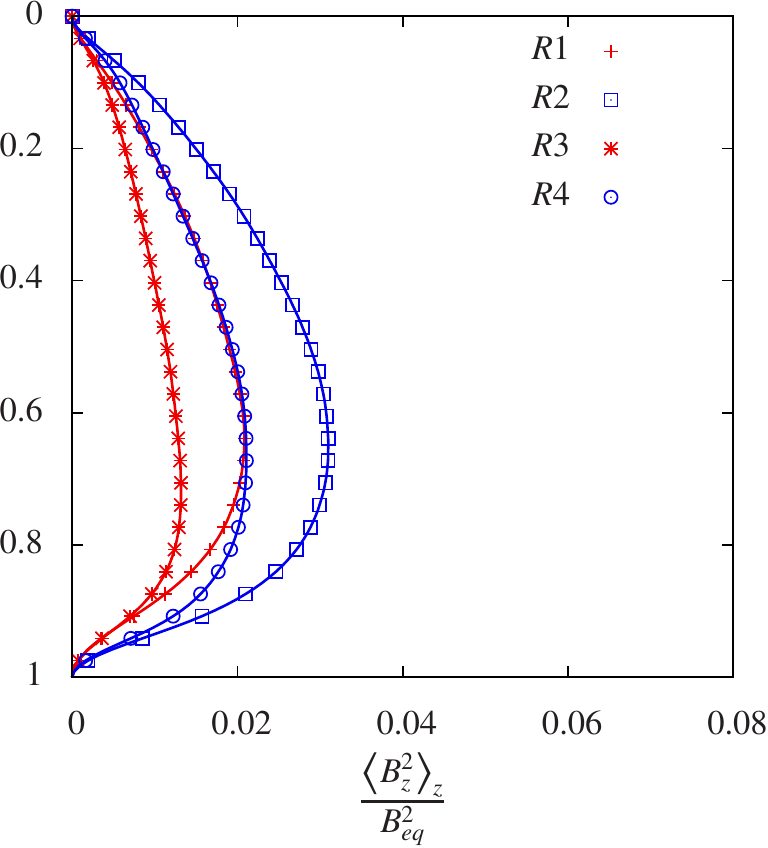}}
        \put(0,-3){\includegraphics[height=130\unitlength]{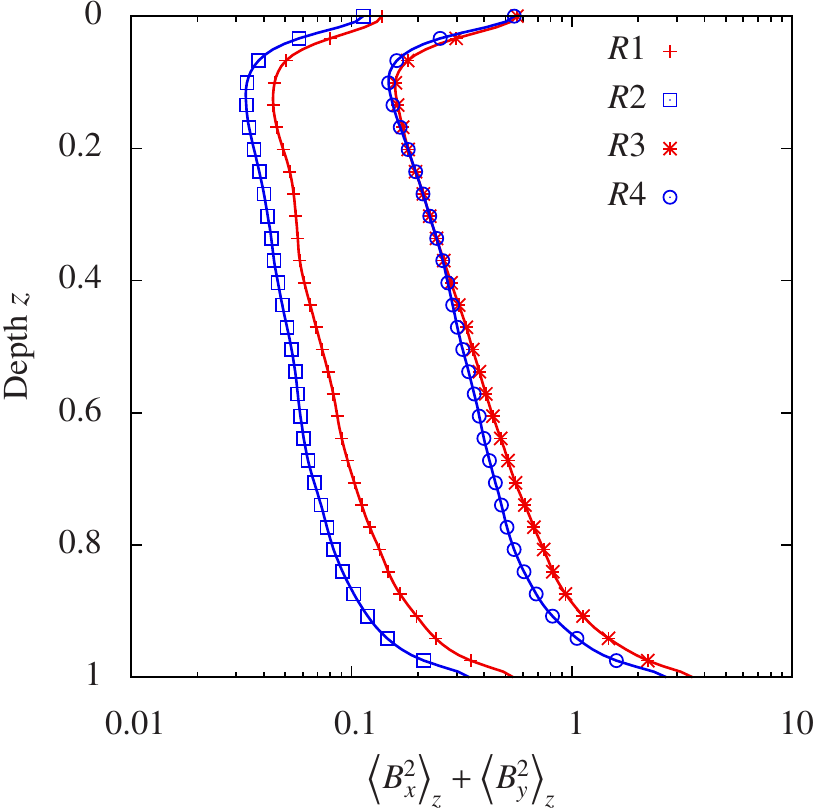}}
        \put(145,-3){\includegraphics[height=130\unitlength]{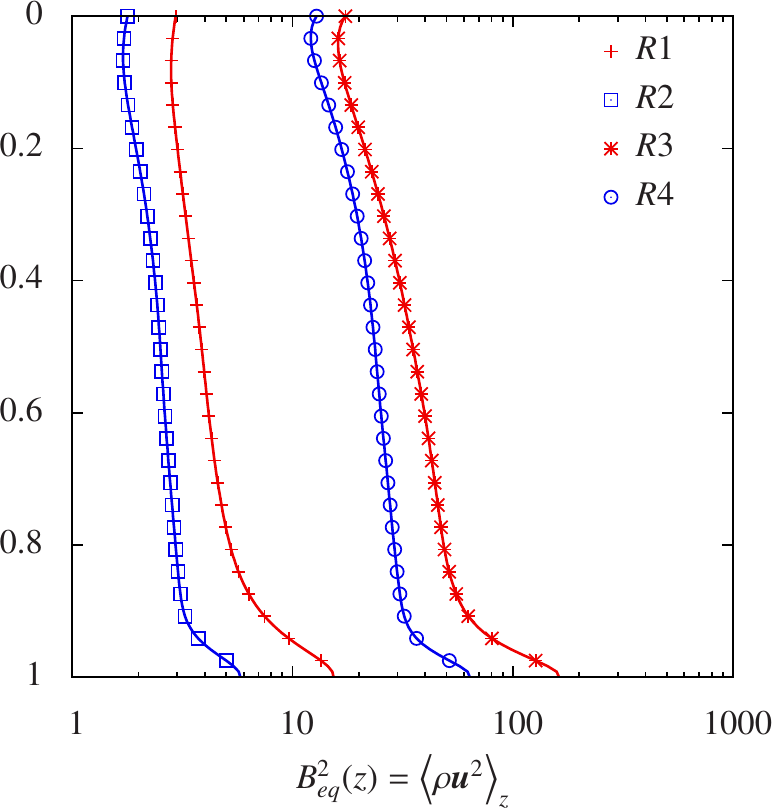}}
	\put(110,160){(a)}
	\put(246,160){(b)}
	\put(114,25){(c)}
	\put(250,25){(d)}
\end{picture}
\caption[]{(a) The horizontal magnetic energy, normalised by the equipartition energy $B_{eq}^2(z)=\left<\rho |\bm{u}|^2\right>_z$. (b) The equivalent plot for the vertical magnetic energy density. (c) The (unnormalised) horizontal magnetic energy density. (d) A plot of $B_{eq}^2(z)$ as a function of depth. Note that figures (c) and (d) are presented in semi-log scale.\label{fig:bn}}
\end{figure}
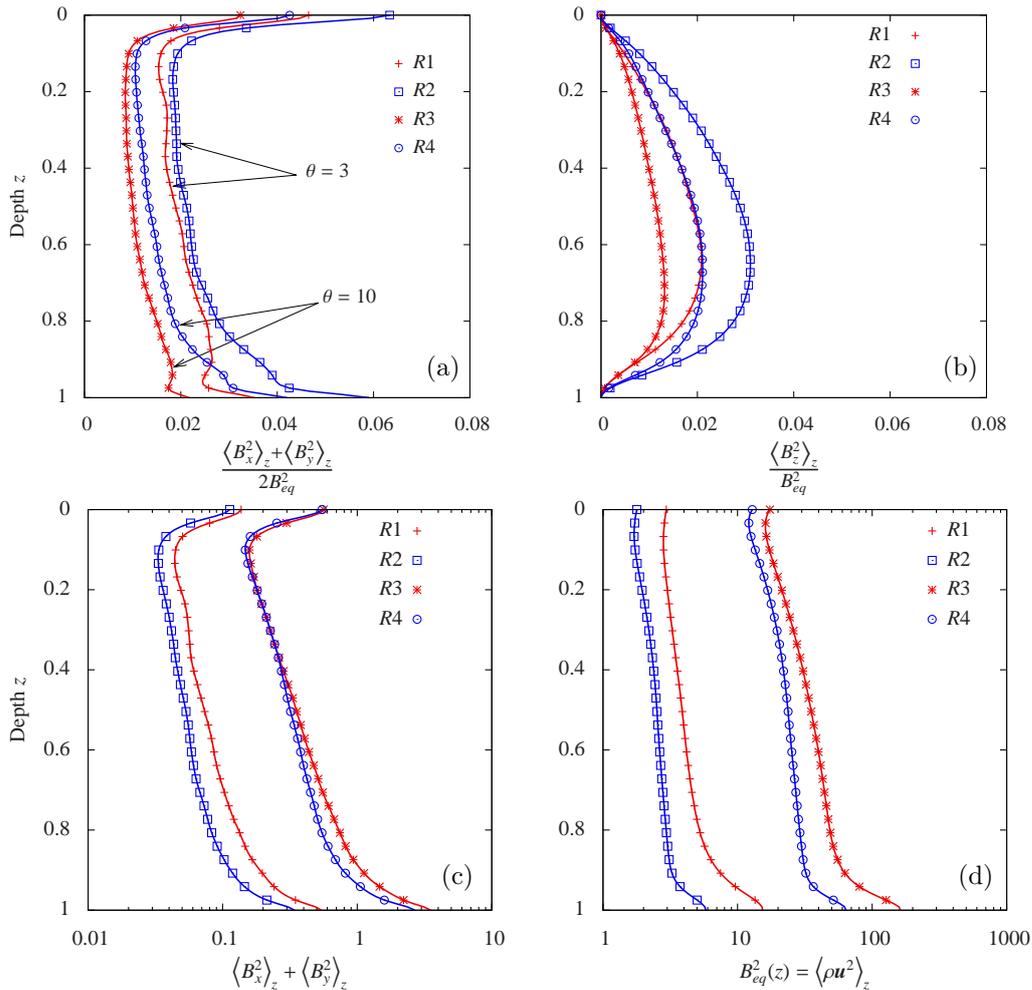

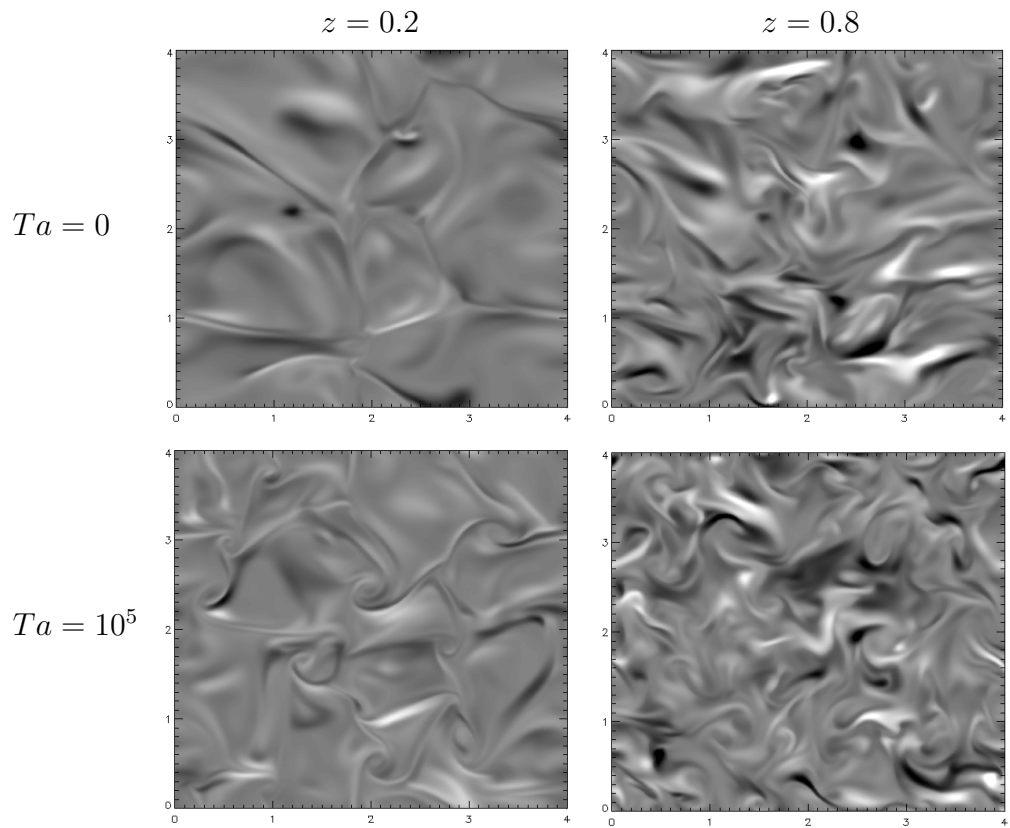
\begin{figure}
\unitlength 0.5mm
\begin{picture}(250,220)
        \put(45,106){\includegraphics[height=100\unitlength]{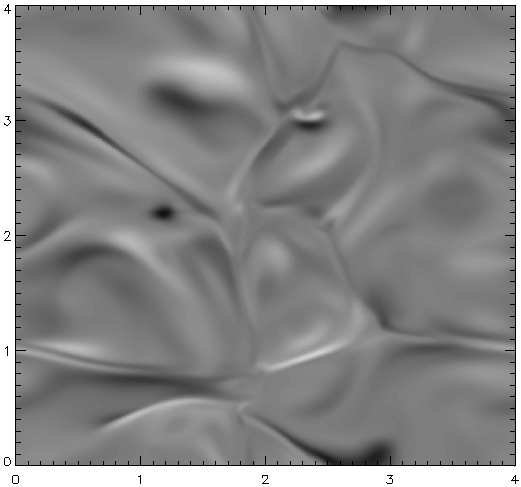}}
        \put(160,106){\includegraphics[height=100\unitlength]{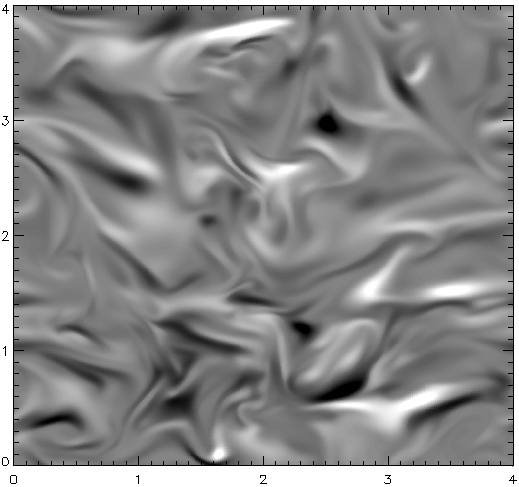}}
        \put(45,-0.2){\includegraphics[height=100\unitlength]{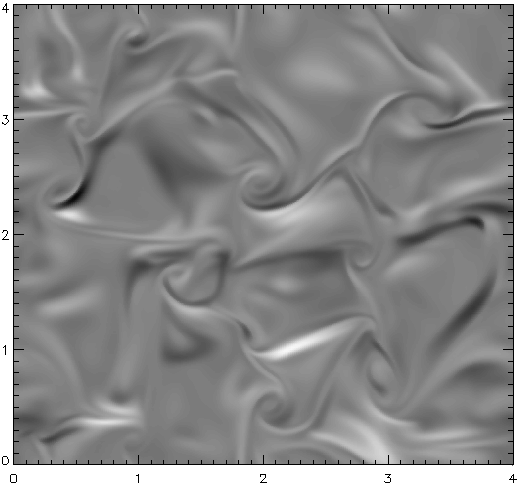}}
        \put(160,-1){\includegraphics[height=100\unitlength]{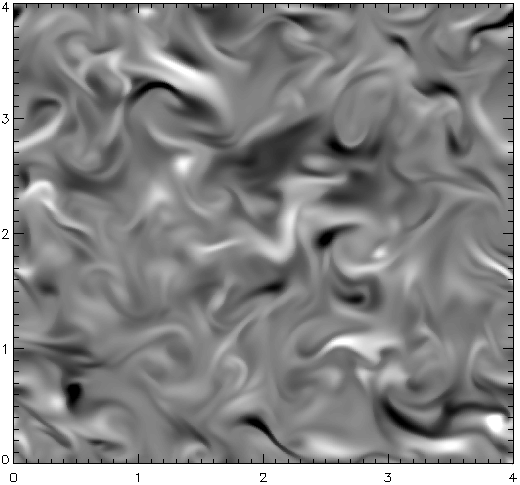}}
	\put(86,210){\large{$z=0.2$}}
	\put(202,210){\large{$z=0.8$}}
	\put(5,156){\large{$Ta=0$}}
	\put(5,50){\large{$Ta=10^5$}}
\end{picture}
\caption[]{Grey-scale plots of the horizontal component $B_x$ of the magnetic field at two different depths, $z=0.2$ and $z=0.8$, for both rotating and non-rotating cases in the saturated phase. The thermal stratification is $\theta=3$ in all cases. Contours are evenly spaced between $B_x=-0.5$ and $B_x=0.5$ for $Ta=0$ and between $B_x=-0.3$ and $B_x=0.3$ for $Ta=10^5$. Light and dark tones correspond to opposite polarities.\label{fig:planform}}
\end{figure}

A more detailed description of the magnetic field in the saturated
phase can be obtained by considering the horizontal and vertical
components of the magnetic energy density. When comparing different
cases, it is useful to normalise the magnetic energy densities by the
local equipartition energy defined by $B_{\textrm{eq}}^2(z)=\left<\rho
  |\bm{u}|^2\right>_z$. The normalised horizontal magnetic energy
density $\left<B_x^2\right>_z+\left<B_y^2\right>_z$ and the normalised
vertical magnetic energy density $\left<B_z^2\right>_z$ are shown in
Figures~\ref{fig:bn}(a) and (b). The unnormalised horizontal magnetic
energy density is shown in Figure~\ref{fig:bn}(c), whilst the depth
dependence of the equipartition energy $B_{\textrm{eq}}^2(z)$ is shown
in Figure~\ref{fig:bn}(d). In the middle of the layer, the horizontal
magnetic energy density is comparable to the vertical magnetic energy
density, whereas it clearly dominates close to the boundaries, as
expected from the chosen boundary conditions. It is clear that the
(unnormalised) horizontal magnetic energy density is stronger at the
lower boundary than it is at the upper boundary, although these are
more comparable when they are scaled in terms of the equipartition
field strength. Note that all of these magnetic energy densities are
sub-equipartition. Although there are some
  quantitative differences in these curves, the variation with depth
  of each of these quantities is broadly similar in most cases. The
  main difference to note is that the vertical component of the
  magnetic energy density tends to be larger in the rotating cases.

Figure~\ref{fig:planform} shows some examples of contours of the
horizontal component of the magnetic field, $B_x$, during the
saturated phase. In all cases, $\theta=3$, and two different depths
are considered ($z=0.2$ and $z=0.8$). The upper row
corresponds to the non-rotating case, $R1$, whereas the lower row
corresponds to the rotating case, $R2$.  It is useful to
compare this figure with the corresponding temperature and enstrophy
plots in Figure~\ref{fig:lyapplot}. Near the upper boundary
(\textit{i.e.} at $z=0.2$), there is a clear correlation between the
convective downflows and the strongest magnetic fields. The
characteristic scale of the magnetic field is smaller in the rotating
case and one observes that the helical structures of the fluid motion
(see Figure~\ref{fig:lyapplot}) affect the magnetic field. Close to
the lower boundary (\textit{i.e.} at $z=0.8$), both cases are similar
even if the characteristic scale of the magnetic structures is still
smaller in the rotating case. Note also that the same colour table is
used for both upper and lower layers. The magnetic field is therefore
stronger and more intermittent near the lower boundary in both cases,
as already illustrated by Figure~\ref{fig:bn}(c). 

\begin{figure}
\unitlength 0.5mm
\begin{picture}(250,150)
        \put(-5,0){\includegraphics[height=140\unitlength]{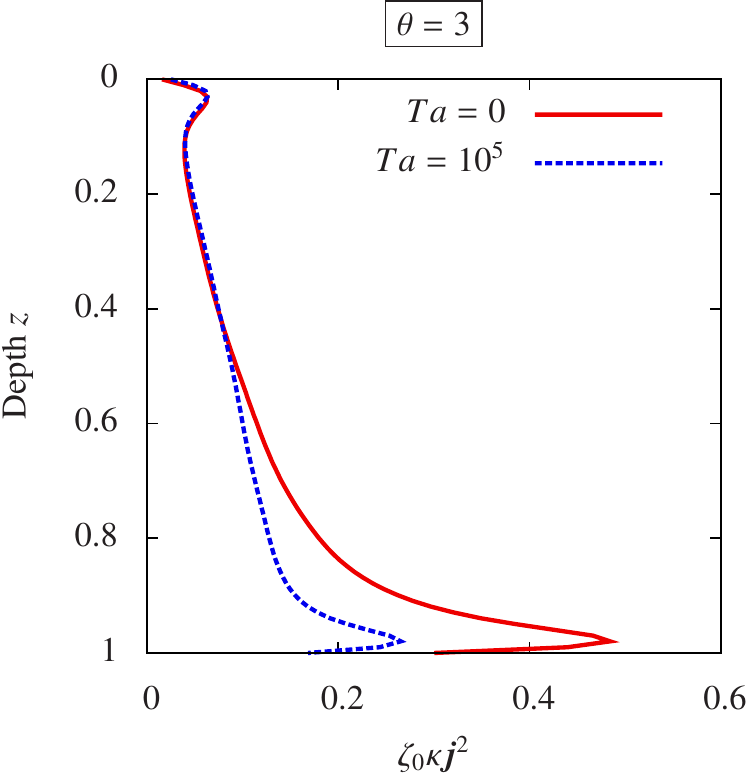}}
        \put(145,0){\includegraphics[height=140\unitlength]{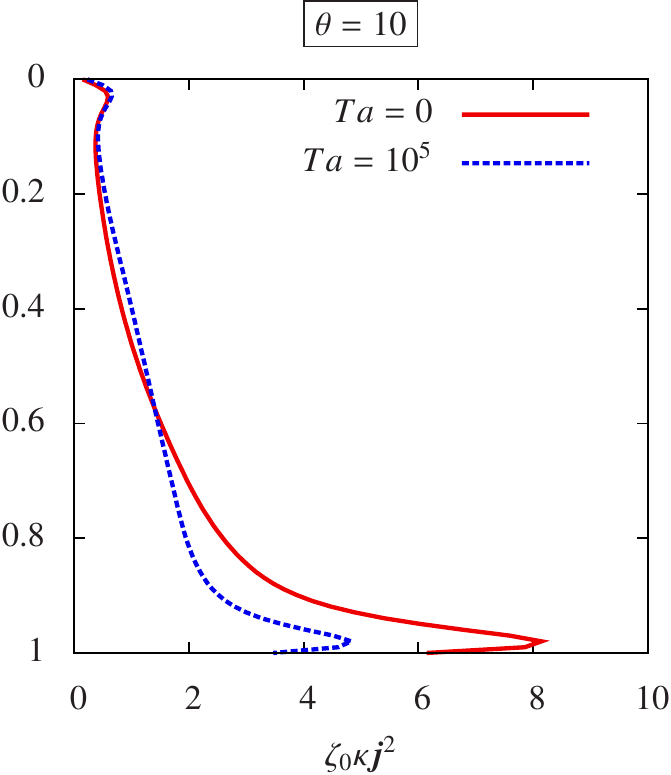}}
	\put(114,26){(a)}
	\put(251,26){(b)}
\end{picture}
\caption[]{The horizontally-averaged ohmic dissipation,
  $\zeta_0\kappa\left<\bm{j}^2\right>_z$, for (a) $\theta=3$ and (b) $\theta=10$.\label{fig:j}}
\end{figure}

The magnetic field geometry plays a crucial role in
  determining the rate at which magnetic energy is dissipated via
  ohmic diffusion. This is clearly an important process to consider
  when comparing the efficiency of convectively-driven dynamos. In
  this model, the rate of magnetic energy dissipation (per unit volume) due to
  ohmic diffusion is given by $\zeta_0\kappa |\bm{j}|^2$, where
  $\bm{j}=\nabla\times\bm{B}$ is the current density. This expression
  for the ohmic dissipation can be derived by taking the scalar
product of the induction equation~\eqref{eq:induction} with $\bm{B}$,
and then integrating over the domain. Note that the
  rate of dissipation is also proportional to the ohmic heating term in
  Equation~\eqref{eq:heateq}. The rate of ohmic dissipation in the
nonlinear phase, averaged
over time and horizontal coordinates, is plotted in
Figure~\ref{fig:j}(a) for $\theta=3$ and Figure~\ref{fig:j}(b) for
$\theta=10$. The dissipation is much larger in the $\theta=10$ case,
since both the kinetic and the magnetic energy densities are greater
by roughly one order of magnitude. For each value of $\theta$, the
magnitude of the dissipation term near the top of the layer is roughly
the same, whether or not the layer is rotating. On the
other hand, rotation clearly reduces the dissipation of magnetic
energy near the lower boundary. Put another way, the
dissipation term is more weakly dependent upon depth in the rotating
cases than it is in the corresponding non-rotating calculations. A similar reduction of the magnetic dissipation by rotation has
already been observed in rotating homogeneous magnetohydrodynamic turbulence by
\citet{favier11}. It is also worth noting that this
  reduction in magnetic dissipation is not a nonlinear effect: it can
  also be observed in the kinematic phase (although it is more
difficult to quantify such a reduction since it becomes necessary to
normalise the dissipation term by an exponentially growing field in
order to take an appropriate time-average). The
  regions of strongest ohmic dissipation in the non-rotating cases
  coincide with the regions of strongest shear. We have already seen,
  in Figures~\ref{fig:lyapplot}(a) and~\ref{fig:lyapplot}(b), that the
  stretching is more homogeneously distributed across the layer in the
  rotating calculations. As a result, the dynamo-generated magnetic
  field organises itself in
such a way that the rate of dissipation in these cases is lower than
it is in the equivalent non-rotating calculations. So
  even though the scaled Lyapunov exponents, as shown in
  Figure~\ref{fig:lyapplot}(b), suggest that there is generally less
  stretching in the rotating cases, this reduction in dissipation
  explains why rotating convection can still act as an efficient
  dynamo (at least at moderate magnetic Reynolds numbers).

\subsection{Saturation mechanisms\label{sec:sat}}

In the previous subsection, we discussed nonlinear
  results, without really saying anything about the saturation
  mechanism for these dynamo calculations. Due to the complexity of
  these dynamo models, it is difficult to say
  anything definitive about this. However, some insights into the
  saturation process can be gained by comparing certain aspects of the
  kinematic and nonlinear phases of the dynamo.

\begin{figure}
\unitlength 0.5mm
\begin{picture}(250,150)
        \put(0,0){\includegraphics[height=140\unitlength]{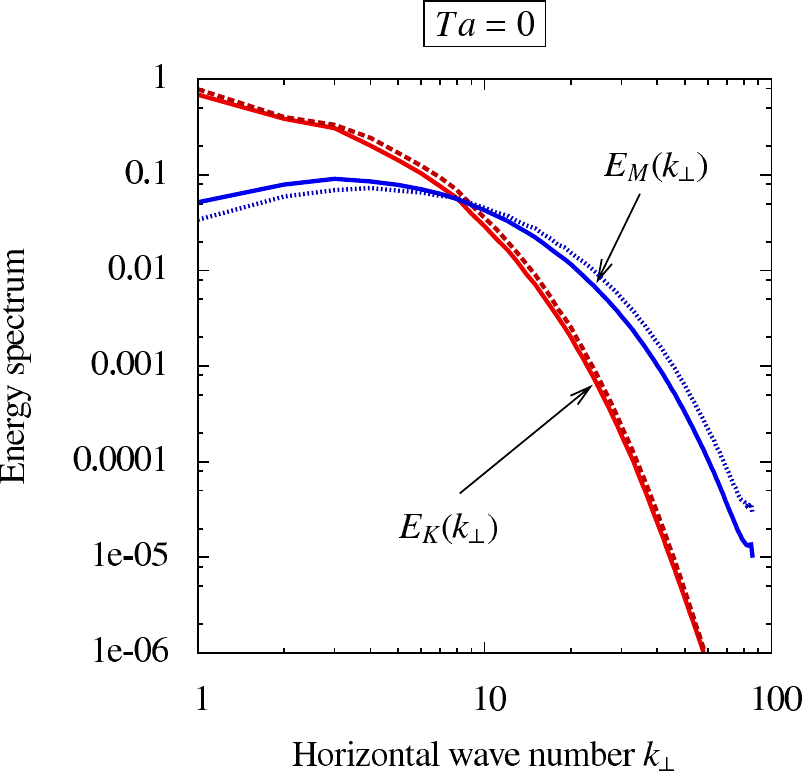}}
        \put(145,0){\includegraphics[height=140\unitlength]{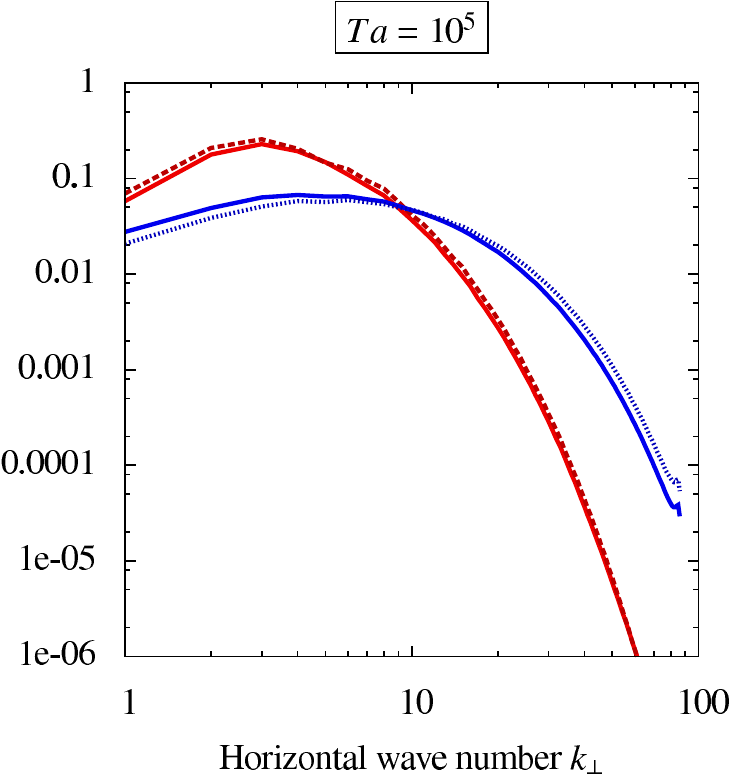}}
	\put(126,130){(a)}
	\put(260,130){(b)}
\end{picture}
\caption[]{Energy spectra for $\theta=3$ (the results
    are similar for $\theta=10$). The solid lines correspond to the
    saturated phase whereas the dotted lines correspond to the
    kinematic phase. The kinetic energy spectrum $E_K(k_{\perp})$ is
    averaged over time and vertical coordinate. The magnetic energy
    spectrum is normalised by the time-average of the magnetic energy
    for the saturated dynamo.\label{fig:spect}}
\end{figure}

Firstly, we consider some of the global properties of the
  dynamo. We define the magnetic energy spectrum in the following way:
\begin{equation}
\label{eq:spectm}
E_M(k_{\perp})=\frac12\sum_{z}\sum_{k_{\perp}} \hat{\bm{B}}(k_x,k_y,z)\cdot\hat{\bm{B}}^*(k_x,k_y,z)
\end{equation}
where $k_{\perp}^2=k_x^2+k_y^2$ is the horizontal wave number,
$\hat{\bm{B}}(k_x,k_y,z)$ is the two-dimensional Fourier transform of
$\bm{B}(x,y,z)$ and the star denotes the complex
conjugation. Similarly, the kinetic energy spectrum is defined as follows:
\begin{equation}
E_K(k_{\perp})\!=\!\frac14\sum_z\!\sum_{k_{\perp}}\!\widehat{\bm{u}}(k_x,k_y,z)\!\cdot\!\widehat{\rho\bm{u}}^*\!(k_x,k_y,z)+\widehat{\bm{u}}^*\!(k_x,k_y,z)\!\cdot\!\widehat{\rho\bm{u}}(k_x,k_y,z) \ .
\end{equation}
In Figure~\ref{fig:spect}, we show the energy spectra corresponding
to the $\theta=3$ calculations ($R1$ and $R2$), in both the kinematic
and nonlinear regimes. The kinetic energy spectrum is time-averaged,
whereas the magnetic energy spectrum is normalised by the
time-average of the magnetic energy for the saturated dynamo before
also being time-averaged. The
results are qualitatively similar for the highly-stratified
case. Comparing the kinetic energy spectra, we see that there is less
kinetic energy at large scales (\textit{i.e.} for $1<k_{\perp}<3$) in
the rotating case. This is connected to the rotationally-induced
reduction in the horizontal scale of motion. Whether or not rotation
is present, the magnetic energy spectra always peak at small scales,
which is consistent with the fact that we do not observe a large-scale
dynamo in these simulations. Comparing the kinematic and
nonlinear phases, we see that the kinetic energy spectra are only
weakly perturbed by the magnetic fields. So there is no evidence here to
suggest that saturation is accompanied by a significant variation in the
kinetic energy spectrum. More interestingly, there is a small
(but perhaps significant) alteration of the magnetic energy spectra as
the dynamo saturates. There is a clear reduction of magnetic energy at
small scales and a corresponding increase at large scales. This is
observed in both the rotating and the non-rotating cases. Although not
shown here, the same trend is observed in the $\theta=10$
calculations. We note that this is consistent with the increase of the
Taylor microscale of the magnetic field reported by
\citet{brandenburg1996}.

\begin{figure}
\unitlength 0.5mm
\begin{picture}(250,150)
        \put(0,0){\includegraphics[height=140\unitlength]{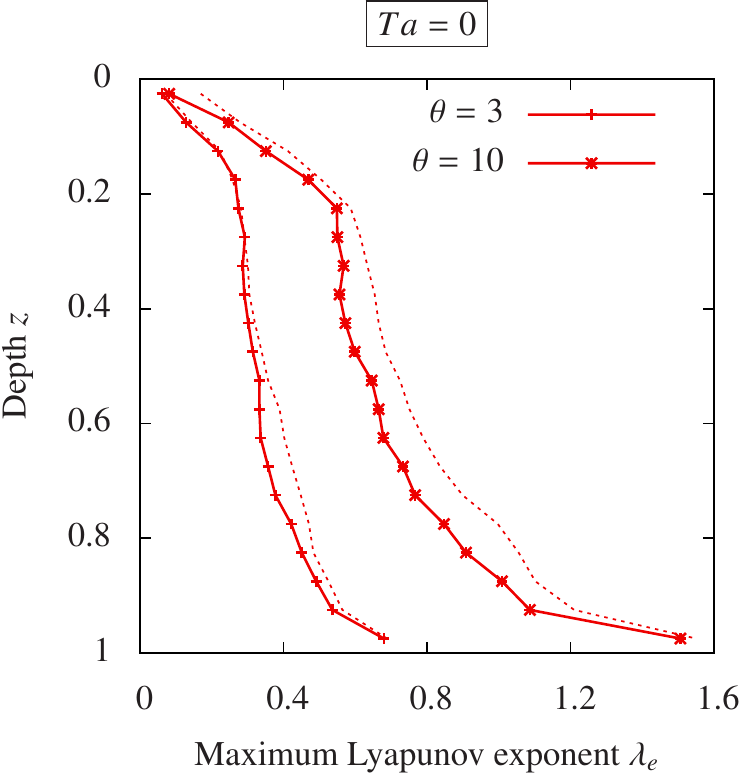}}
        \put(145,0){\includegraphics[height=140\unitlength]{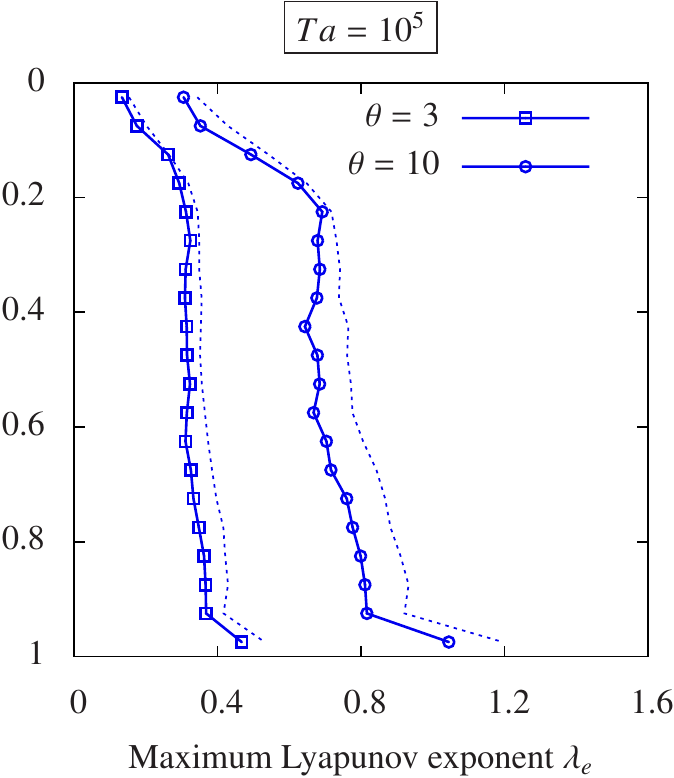}}
	\put(126,130){(a)}
	\put(260,130){(b)}
\end{picture}
\caption[]{Horizontally-averaged maximum Lyapunov exponents versus
  depth during the nonlinear phase (solid lines). The dotted lines
  correspond to the previous kinematic results (see figure
  \ref{fig:lyapplot}(a)).\label{fig:lyapnl}} 
\end{figure}

When considering potential saturation mechanisms,
  one of the most interesting things to consider is whether or not
  the stretching properties of the flow are modified by the magnetic
  fields. Local variations in the stretching would probably not lead
  to significant variations in the kinetic energy spectrum. However,
  we would expect to see changes in the Lyapunov
  exponents if the stretching properties of the flow are modified in
  the nonlinear dynamo regime. Using the flows from our nonlinear
calculations, we evaluate the Lyapunov exponents using exactly the
same procedure as for the kinematic phase (as described in
Section~\ref{sec:lyap}). The particle pairs  are followed from $t=300$
to $t=500$ in the $\theta=3$ cases, and from $t=150$ to $t=250$ in the
$\theta=10$ cases.  The (horizontally-averaged) maximum Lyapunov
exponents for the four cases are shown in Figure~\ref{fig:lyapnl} by
the solid lines. The dotted lines show the corresponding Lyapunov
exponents from the kinematic calculations. In the non-rotating cases,
regardless of the value of $\theta$, the Lyapunov exponents are
slightly lower in the nonlinear phase everywhere except near the upper
and lower boundaries. In the rotating cases, the Lyapunov exponent is
slightly reduced almost everywhere, even close to the lower
boundary. So there is some indication of a
  suppression of stretching due the presence of magnetic
  fields. Furthermore, given the intermittent nature of the magnetic
  field distribution, we might expect the local reduction in stretching
  (in regions of strong magnetic fields) to be greater than that
  suggested by these horizontally-averaged quantities. However,
  this does not mean that the saturation of the dynamo can be explained
  simply by a reduction in stretching. \citet{cattaneo09} have shown
  that a dynamo-saturated velocity field in Boussinesq convection can still
  act as a kinematic dynamo (provided that the new seed field is not
  aligned with the original magnetic field). Although this has not
  been tested here, the depth-dependence of the Lyapunov exponents in
  the nonlinear regime is qualitatively similar to that of the
  original hydrodynamic flows, so we would speculate that these dynamos would
  exhibit similar behaviour. Certainly we do not see the drastic
  reduction in the Lyapunov exponents that \citet{cattaneo96} observed
  in their model of nonlinear dynamo action in a much simpler
  one-scale flow. The saturation process in convectively-driven dynamos
  appears to be more subtle than this.

\begin{figure}
\unitlength 0.5mm
\begin{picture}(250,140)
        \put(-5,0){\includegraphics[height=130\unitlength]{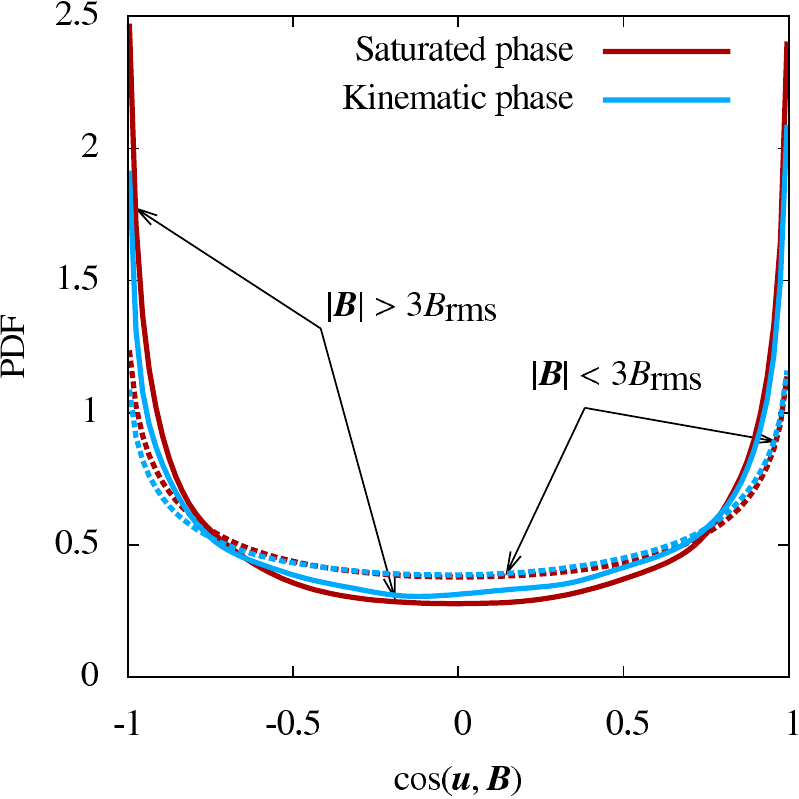}}
        \put(140,0){\includegraphics[height=130\unitlength]{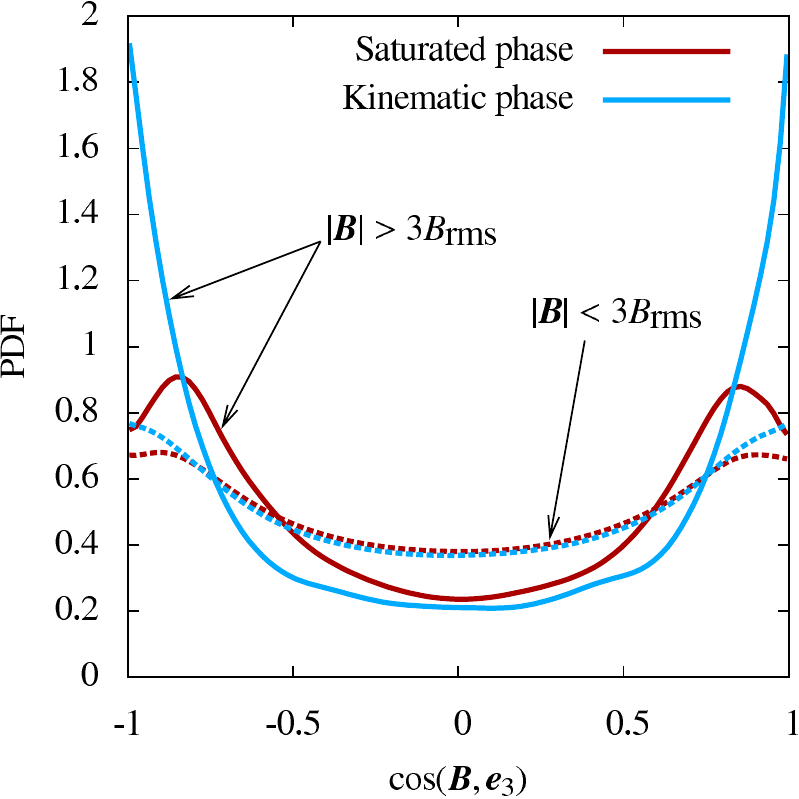}}
	\put(126,130){(a)}
	\put(260,130){(b)}
\end{picture}
\caption[]{Alignment properties for the simulation $R2$ (the results are qualitatively similar for our other simulations). (a) Probability density function of the cosine of the angle between $\bm{u}$ and $\bm{B}$. We separate contributions from points where $|\bm{B}|>3B_{\textrm{rms}}$ (solid lines) and $|\bm{B}|<3B_{\textrm{rms}}$ (dotted lines). (b) Same as (a) but for the angle between $\bm{B}$ and $\bm{e}_3$, where $\bm{e}_3$ is the eigenvector associated to the largest eigenvalue of the rate of strain tensor.\label{fig:cos}}
\end{figure}

We also consider the possibility that the correlation
  between $\bm{u}$ and $\bm{B}$ plays some role in the saturation
  process. The alignment between these two vectors, and the influence
  that this has upon some of the key nonlinearities in the dynamo
  system, has been extensively studied in recent years \citep[see, for
  example,][]{servidio08}. Figure \ref{fig:cos}(a) shows the probability density function of the cosine of the angle between $\bm{u}$ and $\bm{B}$:
\begin{equation}
\cos(\bm{u},\bm{B})=\frac{\bm{u}\cdot\bm{B}}{|\bm{u}||\bm{B}|} \ .
\end{equation}
The pdfs have been obtained by averaging over space (between $z=0.2$ and
$z=0.8$ only, so as to focus upon the interior of the domain) and time
(using around $50$ snapshots in each case). Note that it is not
necessary to carry out any normalisation to compare pdfs in the
kinematic and nonlinear regimes. Given that we would expect any
preferential alignment to be rather localised, we separate regions of
strong magnetic fields from weaker field regions. This is achieved by
plotting two pdfs, one for those mesh points for which
$|\bm{B}|>3B_{\textrm{rms}}$, where $B_{\textrm{rms}}$ is the rms magnetic field, the
other for those mesh points that fall below this threshold field
strength. Despite the fact that this flow is compressible,
inhomogeneous and anisotropic, we see remarkable alignment between
$\bm{u}$ and $\bm{B}$. Perhaps unsurprisingly, the alignment is always
slightly more pronounced for strong magnetic fields than for the
weaker fields. However, the strong alignment between the two fields is
observed in both the kinematic and the nonlinear phases, and is therefore a
property of the induction equation. Hence the saturation of these
convectively-driven dynamos cannot be explained by a modified alignment between
$\bm{u}$ and $\bm{B}$. 

Following \citet{brandenburg1996}, we also consider the alignment between $\bm{B}$ and the eigenvectors of the rate of strain tensor $\bm{\mathsf{S}}$.
The symmetric matrix $S_{ij}=\frac12\left(u_{i,j}+u_{j,i}\right)$ is
computed at each mesh point and the three corresponding eigenvalues
are ordered from the smallest to the largest. The smallest eigenvalue is always
negative and its eigenvector corresponds to the direction of
compression. The magnetic field is found to be mostly perpendicular to
this direction \citep[as already reported by][]{brandenburg1996}, in
both the kinematic and the nonlinear phases. The largest eigenvalue is
always positive and its eigenvector, denoted here by $\bm{e}_3$,
corresponds to the direction of maximum stretching. The probability
density function of $\cos(\bm{B},\bm{e}_3)$, is shown in
Figure~\ref{fig:cos}(b). Note that we again separate the contributions
from the weak field and strong field regions. During the kinematic phase,
the strongest magnetic fields are mostly aligned with the direction of maximum
stretching. Note that there is also some indication of preferential
alignment for the weaker field regions, but this is much less
pronounced. In the saturated phase, the pdf corresponding to the weak
field regions shows a very modest reduction in alignment. However,
there is a dramatic reduction in the alignment between $\bm{B}$ and
$\bm{e}_3$ in the strong field regions. Note that the magnetic fields
for which $|\bm{B}|>3B_{\textrm{rms}}$ only represent $2\%$ of the total number
of points, so that the loss of alignment between the direction of
maximum stretching and the magnetic field is only observed very
locally. This modification of the alignment between $\bm{B}$ and
$\bm{e}_3$ is observed in all our simulations, with or without
rotation, and for both thermal stratifications. This indicates that
this process is rather robust. Given that this modified alignment
will reduce the efficiency of the dynamo in regions of strong magnetic
fields, we conclude that this effect plays a role in the saturation of
convectively-driven dynamos.

%
%

\section{Conclusions\label{sec:conclusions}}

In this paper, we have investigated small-scale dynamo action in
rotating compressible convection. Regardless of the level of
stratification within the domain, both rotating and non-rotating
convective flows can sustain a small-scale dynamo if the magnetic
diffusivity is small enough. Using a mid-layer value for the magnetic
Reynolds number (based upon the integral scale of the turbulence
rather than the layer depth), rotation seems to
reduce the value of the magnetic Reynolds number above which dynamo
action is observed.
Increasing the thermal stratification also reduces the critical value
of the local magnetic Reynolds number in the non-rotating case.
At high values of the magnetic Reynolds number,
the growth rate of the magnetic energy of the dynamo appears to have a
logarithmic dependence upon $R_M$ in the weakly-stratified,
non-rotating simulation. It is more difficult to fit a scaling law in
the other cases, but an $R_M^{1/2}$ scaling may be more appropriate
here. At a given value of the mid-layer (local) magnetic Reynolds number, the
kinematic growth rates are always larger in rotating convection than
they are in the corresponding non-rotating cases.
This dependence upon rotation cannot be attributed solely to the fact that the
  local Reynolds number is smaller in the rotating cases. So we
  conclude that the Coriolis force plays a key role in determining the
  kinematic dynamo properties of rotating convection. At the highest
computationally accessible values of the magnetic Reynolds number, we
find similar levels of saturation in all of our nonlinear calculations
(with the magnetic energy saturating at about $4-9\%$ of the final kinetic
energy). At first sight, this result is slightly
surprising given that the Lyapunov exponents suggest that there is
less stretching in the rotating cases (particularly near the lower
boundary). However, this is compensated by the fact that magnetic
dissipation seems to be much less efficient in the rotating calculations.
It is difficult to say anything definitive about the
  saturation mechanism for these convectively-driven dynamos. However,
  a comparison between the kinematic dynamo regimes and the nonlinear
  saturated states show a slight reduction in the Lyapunov exponents
  in the nonlinear regime (due to a local reduction of the stretching
  properties of the flow). Furthermore, there is some evidence to
  suggest that a reduction in alignment between the strongest magnetic
  fields and the direction of maximum stretching also plays a role in the
  saturation process.

The next natural step in this study is to explore the parametric dependence of this system in more detail. Of course, given the complexity of this numerical model, it is not feasible to conduct a complete survey of parameter space, although any further calculations that can be carried out will obviously enhance our understanding of this problem. We also intend to investigate the ability of rotating compressible convection to produce a magnetic field on much larger scales than the scales that are associated with the convective motions. The question of how hydromagnetic dynamos are able to generate large-scale magnetic fields is undoubtedly one of the most challenging issues in modern dynamo theory. In the standard formulation of mean-field
dynamo theory \citep{moff78}, this regenerative process relies upon
the presence of helical motions, which are invariably
  produced when a flow is influenced by rotation. Hence, rotating convection should be able to drive a
large-scale dynamo. However, despite predictions from mean-field dynamo theory,
\citet{cattaneo06} failed to find evidence for a large-scale dynamo in
their Boussinesq model. Interestingly, \citet{kapyla2009} did find
large-scale magnetic fields in their compressible model. They argue
that the absence of large-scale fields in the Boussinesq model of
Cattaneo and Hughes can be attributed to a rotation rate that is too
slow. This is a possibility, however the effects of compressibility
may also be playing a role. The flow is (locally) strongly helical in
the Boussinesq calculations of \citet{cattaneo06}, but the mid-layer
symmetry of their setup implies that the mean helicity distribution is
antisymmetric about the mid-plane. It may be important to break this
symmetry in order to generate large-scale fields. In work that
is currently in progress,  we are investigating this issue by carrying
out simulations of rapidly-rotating convection in a wide compressible
layer. The size of the layer is important, because it is necessary to
have a clear separation in scales between the small-scale fields and
any large scale magnetic fields that may be generated. The
computational domain that was considered in the present study was too small
to allow for this separation in scales, which may explain why no large
scale fields were observed here (despite the fact that the helicity of the flow is asymmetric about the mid-plane in these stratified rotating calculations). Of course, other physical ingredients
could also be included into this system once the basic ingredients of
rotation and stratification are properly understood. For example, the
role of the penetrative layer in the model of \citet{kapyla2009} is
also unclear. This may be promoting the formation of large-scale
fields in some way. It may also be of interest to consider the effect
of including a shear flow \citep[see, for example,][]{hughes2009} in this
compressible model.  

{\bf Acknowledgements} The authors wish to thank the anonymous referees
for their helpful comments and suggestions, which improved the quality of the manuscript. B.F. wishes to thank A.W. Baggaley for helpful discussions. This work has been supported by the Engineering and Physical Sciences Research Council through a research grant (EP/H006842/1). All numerical calculations have been carried out using the HECToR supercomputing facility. 

%
%

\bibliographystyle{jfm}
\bibliography{biblio}

\end{document}